\ifdefined\pdfoutput\pdfoutput=1\fi  % force PDFLaTeX on arXiv (first 5 lines)
\documentclass{article}
\usepackage{spconf,amsmath,graphicx,hyperref}
\usepackage{adjustbox}
\usepackage{algorithm}
\usepackage{algorithmic}
\usepackage{multirow}
\usepackage{booktabs}
\usepackage{array}
\usepackage{arydshln}
\usepackage{listings}
\usepackage[table]{xcolor}
\usepackage{amssymb}
\usepackage{threeparttable}
\usepackage{caption}
\usepackage{textcomp}
\usepackage{cuted}
\usepackage{placeins}
\usepackage{dsfont}

\newcolumntype{N}{>{\centering\arraybackslash}p{3.45em}}

\title{TAG-Bench: Benchmarking Temporal Audio Grounding in Large Audio Language Models}

\name{Yuhang Dai$^{\dag}$, Xin Shu$^{\dag}$, Zengxi Li, Lei Xie, Xiangang Li, Jianwei Yu \thanks{$^{\dag}$ Equal contribution.}}
\address{%
  \raisebox{-0.7ex}{%
    \includegraphics[height=3ex]{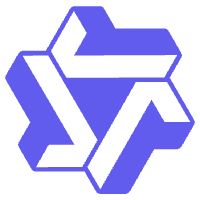}%
  }~%
  \textbf{Alibaba Token Foundry}%
}
\begin{document}
\ninept
\maketitle
\begin{abstract}
Large audio language models (LALMs) can describe \emph{what} is heard, but their ability to localize \emph{when} queried content occurs remains less systematically evaluated. We present TAG-Bench, a benchmark for temporal audio grounding in which a model returns every time interval that matches a natural-language query. TAG-Bench contains 1,750 human-verified query–recording pairs covering 149.5 hours, with eight source-dependent subsets spanning query categories and audio durations from 7 s to 20 min; 22.1\% of the queries have multiple ground-truth intervals. Across 21 evaluated systems, the best-performing model achieves 31.2 mIoU and is the only system above 20 mIoU on the two long subsets, yet even this top performer reaches only 21.5\% recall at IoU$\geq$0.7. Moreover, 9 of 21 systems fall below 5 mIoU, and every model under-reports the number of occurrences on one-to-many queries, with none exceeding 13.2\% count accuracy. Because responses are free-form, we report parsing-failure rate and MAE coverage: parsing failures remain in mIoU, Recall, gIoU, and count metrics as empty predictions but do not enter MAE. The results separate precise localization, occurrence enumeration, and output-format reliability within a benchmark whose cross-subset comparisons are descriptive rather than controlled estimates of query abstraction or duration. We will release the TAG-Bench data and evaluation code to support future research.

\end{abstract}
\vspace{-0.2cm}
\section{Introduction}
\vspace{-0.1cm}
\label{sec:intro}

Large audio language models (LALMs) have advanced rapidly across audio captioning, open-ended audio question answering, and multi-task reasoning benchmarks such as MMAU~\cite{mmau}, MMAU-Pro~\cite{mmaupro}, and MMAR~\cite{mmar}. Audio-language systems such as the Audio Flamingo series~\cite{af,af2,af3}, Kimi-Audio~\cite{kimiaudio}, MiMo-Audio~\cite{mimoaudio}, Step-Audio~2~\cite{stepaudio2}, and MOSS-Audio~\cite{mossaudio} can describe complex acoustic scenes in fluent language. Their published evaluations emphasize audio understanding, reasoning, captioning, or spoken interaction rather than explicit timestamp localization. The complementary question, \emph{when} an event happens, is a prerequisite for many practical applications, including meeting and podcast navigation, audio forensics and content moderation, acoustic surveillance, and audio editing. A model that can state that a dog barked but cannot indicate the moment it barked is of limited use to a user who needs to jump to that moment in an hour-long recording.

\begin{figure}[t]
  \centering
  \includegraphics[clip, trim=0cm 7cm 0cm 0cm, width=\columnwidth]{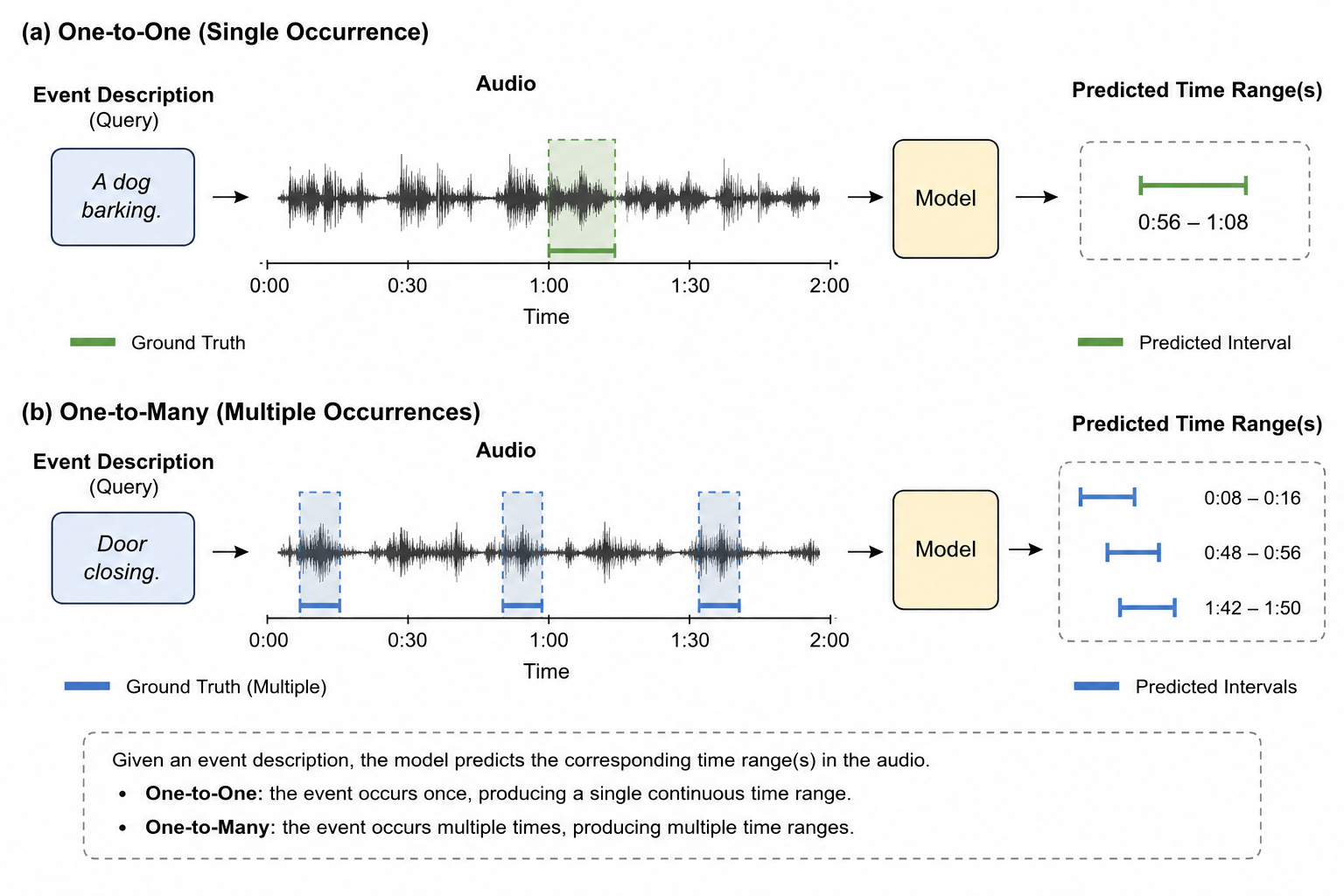}%左下右上
  \vspace{-0.2cm}
  \caption{Examples of TAG tasks. \textbf{(a) Single-interval:} a query event occurs once, and the model predicts its single temporal boundary. \textbf{(b) Multi-interval:} a query event occurs multiple times, and the model predicts all corresponding time intervals.}
  \label{fig:tag_des}
  \vspace{-0.5cm}
\end{figure}

We refer to this capability as \emph{\textbf{T}emporal \textbf{A}udio \textbf{G}rounding} (TAG): given an audio recording and a natural-language query, the model must return the start and end times of every segment whose content matches the query. TAG is challenging for current LALMs for at least three reasons. First, the target is usually a needle in a haystack: in realistic recordings, the relevant segment covers only a small fraction of the audio (a median of 7\% in our benchmark). Second, a query may match several disjoint segments, for example recurring footsteps or a repeatedly discussed topic, so the model must recall all occurrences; this one-to-many setting has only recently been studied even in the video domain~\cite{omtg}. Third, current autoregressive multimodal decoders remain inaccurate at precise timestamp prediction, a limitation also observed in video temporal grounding~\cite{timelens,timemarker,distime}. More broadly, text-only evaluations also show persistent limitations in temporal reasoning~\cite{timebench}.

Existing audio resources cover parts of this problem but not the whole. Text-to-audio grounding~\cite{t2ag} and temporally aligned resources such as AudioTime~\cite{audiotime} and TACOS~\cite{tacos} provide temporal supervision on short clips using event-level or temporally aligned free-text descriptions. The DCASE audio moment retrieval task targets query-based retrieval with dedicated detection architectures~\cite{dcase_kim,dcase_sugawara} rather than broadly trained LALMs. Recent time-aware LALM efforts, including TimeAudio~\cite{timeaudio}, LAT-Audio~\cite{lataudio}, GigaChat Audio~\cite{gigachat}, and audio-side time-prompt post-training~\cite{timepro}, report temporal results under differing datasets and protocols. Multi-event~\cite{sensetag} further reports degraded detection and more false alarms as the number of events increases. To our knowledge, prior resources do not jointly cover all four properties targeted by TAG-Bench: (i)~queries from a single word to abstract semantics, (ii)~audio from seconds to tens of minutes, (iii)~one-to-many targets, and (iv)~evaluation of heterogeneous LALMs under one protocol.

TAG-Bench fills this gap. Our contributions are as follows.
\begin{itemize}
\setlength\itemsep{0.1em}
\item \textbf{A source-diverse, multi-granularity TAG benchmark.} 1{,}750 human-verified queries over 1{,}750 distinct recordings (149.5 hours), organized into eight diagnostic subsets spanning words, acoustic events and descriptions, semantics, emotion, and audio durations from 7\,s to 20\,min.
\item \textbf{One-to-many grounding with dedicated diagnostics.} 22.1\% of the queries carry multiple ground-truth intervals. Beyond a union-based IoU that rewards recalling every occurrence, we introduce a count-accuracy diagnostic that decouples ``finding all occurrences'' from ``placing boundaries precisely.''
\item \textbf{A unified evaluation of 21 systems}, spanning general-
purpose large audio understanding LLMS, omni-modal systems (including the commercial Gemini-3.1-Pro), and temporal-grounding expert models. Subset-level, one-to-one versus one-to-many, parsing-coverage, and model-series analyses separate boundary error, missed occurrences, and output-format failure.
\end{itemize}

TAG-Bench is a diagnostic suite rather than a controlled factorial experiment: query category, source domain, audio duration, and recurrence rate are not varied independently. We therefore treat cross-subset contrasts as descriptive and reserve causal claims about abstraction or context length for matched-data studies; Sec.~\ref{ssec:scope} states these boundaries explicitly.

\begin{figure*}[ht]
  \centering
  \includegraphics[clip, trim=0cm 0cm 0cm 0cm, width=0.92\textwidth]{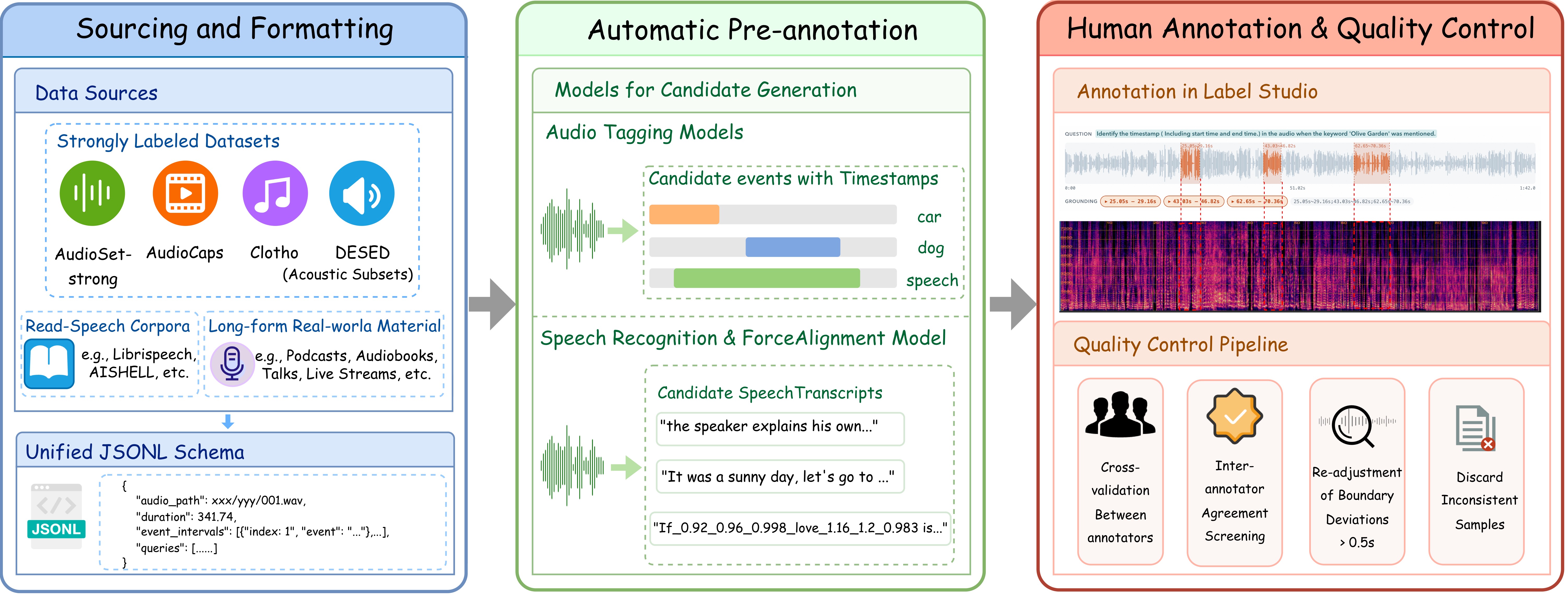}
  \caption{The TAG-Bench construction pipeline. (1)~\emph{Sourcing and formatting}: strongly labeled datasets, read-speech corpora, and long-form real-world material are normalized into a unified JSONL schema. (2)~\emph{Automatic pre-annotation}: audio tagging (PANNs) and speech recognition (Whisper) models generate candidate timestamps and transcripts. (3)~\emph{Human annotation and quality control}: annotators refine the candidates in Label Studio with waveform and spectrogram views, followed by cross-validation, agreement screening, and re-adjudication of large boundary deviations.}
  \label{fig:pipeline}
  \vspace{-0.5cm}
\end{figure*}

\vspace{-0.2cm}
\section{Related Work}
\vspace{-0.1cm}
\label{sec:related}

\textbf{Audio understanding benchmarks.} MMAU~\cite{mmau}, MMAU-Pro~\cite{mmaupro}, and MMAR~\cite{mmar} assess multi-task audio reasoning through question answering; their published task descriptions emphasize question answering and reasoning rather than timestamp-output evaluation. Text-to-audio grounding was formalized on AudioCaps-style clips~\cite{t2ag}. AudioTime~\cite{audiotime} and TACOS~\cite{tacos} provide temporally aligned data on clips no longer than 30 seconds, principally for temporal-control or language--audio representation research. TAG-Bench differs in query diversity (covering speech content, semantics, and emotion in addition to acoustic events), in audio length (up to 20 minutes), and in its explicit one-to-many setting.

\textbf{Temporal grounding in video and text.} Recent video temporal-grounding work includes MLLM-based grounding baselines and benchmarks~\cite{timelens,timelens2,timemarker}, timestamped dense-captioning systems~\cite{timechatcap}, distribution-based time representations~\cite{distime}, and the one-to-many formulation~\cite{omtg}. TimeBench~\cite{timebench} shows that even text-only temporal reasoning remains difficult for LLMs. These findings motivate a dedicated audio counterpart, in which the signal offers no visual anchors and events can be short, overlapping, and acoustically subtle.

\textbf{Time-aware audio LLMs.} A growing line of work injects temporal awareness into LALMs: TimeAudio~\cite{timeaudio} bridges temporal gaps between frames, LAT-Audio~\cite{lataudio} uses progressive global-to-local reasoning with iterative local-audio tool calls, GigaChat Audio~\cite{gigachat} answers with explicit timestamps over long recordings, audio-side time prompts are used in post-training~\cite{timepro}, and retrieval-based pipelines address event-grounded question answering over hours of audio~\cite{larag}. DCASE Task~6 systems~\cite{dcase_kim,dcase_sugawara} and WhisperX~\cite{whisperx} provide dedicated methods for audio-moment retrieval and word-level timestamp alignment, respectively. TAG-Bench provides a common evaluation setting for comparing general and specialist temporal-grounding approaches.

\vspace{-0.2cm}
\section{TAG-Bench}
\vspace{-0.1cm}
\label{sec:bench}
\vspace{-0.2cm}
\subsection{Task Definition}
\vspace{-0.1cm}
Given an audio recording $A$ and a natural-language query $q$, the model must output a set of time intervals $\hat{Y}=\{(\hat{s}_i,\hat{e}_i)\}$ covering all segments of $A$ whose content matches $q$. The ground truth $Y$ is likewise a set of one or more intervals. One-to-one and one-to-many queries share the same output format.
% (e.g., \texttt{12.4s\textasciitilde15.9s;42.0s\textasciitilde44.2s}).
\vspace{-0.2cm}
\subsection{Benchmark Composition}
\vspace{-0.1cm}
TAG-Bench contains 1{,}750 queries, each paired with a distinct audio recording, for a total of 149.5 hours of audio. The queries are organized into eight subsets along two dimensions, query category and audio length, as summarized in Table~\ref{tab:stats}. The evaluation unit is a query--recording pair, not an hour of audio: the two long-audio subsets contain 440 query-level outcomes despite accounting for 127.6 hours. Because each recording contributes exactly one query, TAG-Bench does not measure within-recording consistency across different queries.

\begin{table}[t]
\centering
\caption{TAG-Bench composition. \emph{Multi\%} denotes the percentage of queries with $\geq$2 ground-truth intervals; \emph{Avg.\ dur.} is the average audio length.}
\vspace{-0.2cm}
\label{tab:stats}
\setlength{\tabcolsep}{3.2pt}
\renewcommand{\arraystretch}{0.9} % 增加行间距为0.8倍
\resizebox{\columnwidth}{!}{%
\begin{tabular}{lrrrr}
\toprule
Subset & \#Queries & Avg.\ dur. & Total & Multi\% \\
\midrule
Word                        & 279 & 7.2\,s    & 0.56\,h  & 1.4 \\
Acoustic event              & 301 & 60.0\,s   & 5.02\,h  & 41.9 \\
Acoustic description        & 260 & 60.0\,s   & 4.33\,h  & 33.1 \\
Semantic                    & 180 & 91.0\,s   & 4.55\,h  & 4.4 \\
Rough semantic (fuzzy)      & 161 & 93.2\,s   & 4.17\,h  & 16.1 \\
Emotion                     & 129 & 93.2\,s   & 3.34\,h  & 34.9 \\
Long -- semantic            & 180 & 17.3\,min & 51.80\,h & 3.3 \\
Long -- acoustic desc.      & 260 & 17.5\,min & 75.77\,h & 33.1 \\
\midrule
Total                       & 1{,}750 & 307.6\,s & 149.5\,h & 22.1 \\
\bottomrule
\end{tabular}}
\vspace{-0.5cm}
\end{table}

\begin{figure}[t]
  \centering
  \includegraphics[width=\columnwidth]{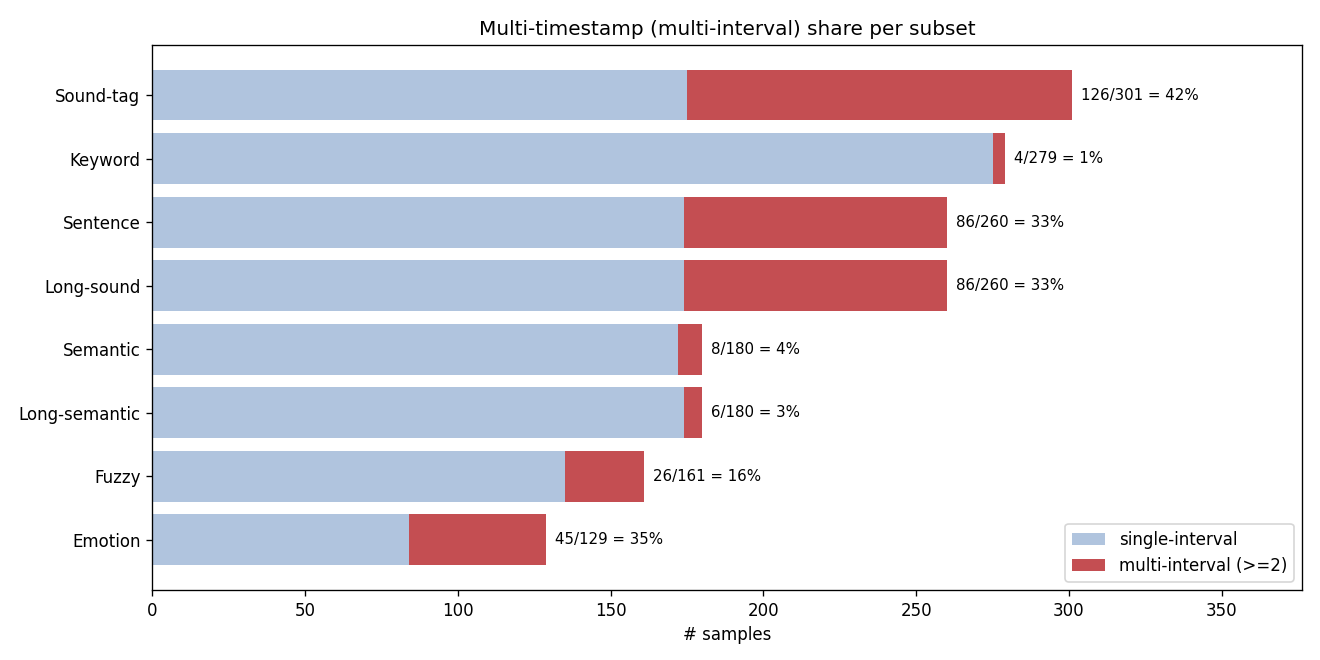}
  \vspace{-0.5cm}
  \caption{One-to-one vs.\ one-to-many queries per subset. Acoustic-event, emotion, and description queries have the largest recurring-target shares, whereas word and semantic queries are mostly one-to-one.}
  \label{fig:multishare}
    \vspace{-0.5cm}
\end{figure}

\textbf{Query categories and source domains.} \emph{Word} queries ask for the position of a single spoken word (median query length of 6 characters). \emph{Acoustic event} queries name a sound class (e.g., birds chirping or an engine), whereas \emph{acoustic description} queries use a full descriptive sentence. \emph{Semantic} queries describe spoken content at the discourse level (median length of 141 characters), \emph{rough semantic} queries use vague or paraphrased wording, and \emph{emotion} queries ask where a specified emotional expression occurs. The word subset is drawn from read speech, the acoustic subsets primarily cover everyday environmental sounds, and the descriptive and semantic subsets include podcast, interview, sports, and finance material. These categories are therefore diagnostic strata, not a controlled abstraction ladder: source domain, duration, and multi-target prevalence also vary, so a cross-subset score gap cannot be attributed to query abstraction alone.

\textbf{Audio-length dimension.} Six subsets use short-to-medium recordings (7\,s--93\,s on average); the two \emph{long} subsets pose the same semantic and acoustic-description queries over recordings averaging about 17 minutes (up to 20 minutes). The two long subsets account for 25\% of the queries but 85\% of the total audio duration, making long-context temporal perception a first-class citizen of the benchmark.

\textbf{One-to-many targets.} 387 queries (22.1\%) carry $\geq$2 ground-truth intervals (up to 18 per query), because sound events such as footsteps or bird calls naturally recur; averaged over these queries, the ground truth contains 4.02 intervals. Fig.~\ref{fig:multishare} shows that the one-to-many share varies strongly across subsets, from 1\% (word) to 42\% (acoustic event). A correct answer must recall \emph{all} occurrences; this setting parallels one-to-many video grounding~\cite{omtg} and is scored with the union-based IoU of Sec.~\ref{ssec:metrics}.
\vspace{-0.2cm}
\subsection{Construction Pipeline}
\vspace{-0.1cm}
TAG-Bench is built in three stages. As shown in Fig.~\ref{fig:pipeline}.(1)~\emph{Sourcing and formatting}: recordings are drawn from strongly labeled open datasets (e.g., AudioSet-strong, AudioCaps, Clotho, and DESED for the acoustic subsets, and read-speech corpora for the word subset) and from long-form real-world material such as podcasts; every item is normalized into a unified JSONL schema with audio path, duration, event intervals, and language queries. (2)~\emph{Automatic pre-annotation}: audio tagging and speech recognition models generate candidate timestamps and transcripts for subsequent human review. (3)~\emph{Human annotation and quality control}: annotators verify and refine the candidate intervals in a customized Label Studio interface with waveform and spectrogram views; annotations then undergo cross-validation between annotators, inter-annotator agreement screening, and re-adjudication of boundary deviations exceeding 0.5\,s, and inconsistent samples are discarded.

\vspace{-0.2cm}
\section{Experimental Setup}
\vspace{-0.1cm}
\label{sec:exp}
\vspace{-0.2cm}
\subsection{Evaluated Models}
\vspace{-0.1cm}
\label{ssec:models}
We use three operational reporting groups rather than mutually exclusive capability classes: (i)~\emph{general-purpose audio LLMs} including Audio Flamingo 2, its Sound-CoT variant, and Audio Flamingo 3~\cite{af2,af2cot,af3}, FireRedAudio-9B~\cite{fireredaudio}, GLM-4-Voice-9B~\cite{glm4voice}, Kimi-Audio-7B~\cite{kimiaudio}, MiDashengLM-7B~\cite{midashenglm}, MiMo-Audio-7B-Instruct~\cite{mimoaudio}, the MOSS-Audio 4B/8B family in instruct and thinking variants~\cite{mossaudio}, the Step-Audio family (Step-Audio-2-mini-Base, Step-Audio-2-mini, Step-Audio-2-mini-Think~\cite{stepaudio2}, and Step-Audio-R1~\cite{stepaudior1}), and LLaMA-Omni2-14B~\cite{llamaomni2}; (ii)~\emph{omni-modal baselines}, comprising LLaMA-3.1-8B-Omni~\cite{llamaomni} and Gemini-3.1-Pro~\cite{gemini31}, with the latter accessed through a commercial API; and (iii)~\emph{temporal-awareness expert models}, namely LAT-Audio~\cite{lataudio}, which uses progressive global-to-local reasoning with iterative local-audio tool calls, and TimeAudio~\cite{timeaudio}, which is post-trained for timestamp prediction. In total, we benchmark 21 systems across these groups. Several families provide thinking and non-thinking variants, enabling within-family comparisons of the two inference paradigms. Every system receives the same item-specific benchmark question text through its model-native chat and audio wrapper. All models receive the same query templates asking for explicit start--end timestamps of every matching segment.

\begin{table*}[t!]
\centering
\caption{Main results on TAG-Bench. \emph{One-to-one}/\emph{One-to-many} are mIoU on queries with a single ground-truth interval (1{,}363 queries) and with multiple intervals (387 queries), respectively. \emph{Fail} is the percentage of responses without a recoverable timestamp (lower is better). All values are percentages. Operational groups are General-purpose Audio Large Language Models (GALLMs.), omni-modal interface (Omni; including a commercial API), and temporal-awareness expert models. Best and second-best grounding scores are in \textbf{bold} and \underline{underlined}; failure rates are not ranked.}
\vspace{-0.2cm}
\label{tab:main}
\setlength{\tabcolsep}{8pt}
\renewcommand{\arraystretch}{0.86}
\begin{adjustbox}{max width=\textwidth}
\begin{tabular}{c|lccccccc}
\toprule
Type & Model & mIoU & R@0.3 & R@0.5 & R@0.7 & One-to-one & One-to-many & Fail \\
\midrule
\multirow{17}{*}{GALLMs.} & Audio Flamingo 2~\cite{af2} & 0.5 & 0.7 & 0.4 & 0.1 & 0.6 & 0.1 & 0.1 \\
 & AF2 Sound-CoT~\cite{af2cot} & 2.9 & 3.4 & 0.9 & 0.2 & 2.8 & 3.3 & 0.1 \\
 & Audio Flamingo 3~\cite{af3} & 8.7 & 12.1 & 5.9 & 2.0 & 8.9 & 7.9 & 0.1 \\
 & FireRedAudio~\cite{fireredaudio} & \textbf{31.2} & \textbf{39.7} & \textbf{30.8} & \textbf{21.5} & \textbf{34.1} & 21.0 & 15.1 \\
 & GLM-4-Voice-9B~\cite{glm4voice} & 3.4 & 4.1 & 1.4 & 0.3 & 3.2 & 4.1 & 33.5 \\
 & Kimi-Audio-7B~\cite{kimiaudio} & 1.4 & 1.4 & 0.5 & 0.2 & 1.5 & 0.9 & 26.9 \\
 & MiDashengLM-7B~\cite{midashenglm} & 4.6 & 6.2 & 1.2 & 0.2 & 4.2 & 5.8 & 28.7 \\
 & MiMo-Audio-7B-Instruct~\cite{mimoaudio} & 12.2 & 17.5 & 9.0 & 3.4 & 12.8 & 9.9 & 1.9 \\
 & MOSS-Audio-4B-Instruct~\cite{mossaudio} & 14.6 & 19.7 & 10.3 & 5.9 & 15.3 & 12.2 & 3.5 \\
 & MOSS-Audio-4B-Thinking~\cite{mossaudio} & 27.8 & 36.3 & 24.1 & 16.2 & 29.5 & \underline{22.0} & 0.9 \\
 & MOSS-Audio-8B-Instruct~\cite{mossaudio} & 12.8 & 17.1 & 9.5 & 5.3 & 13.7 & 9.4 & 11.2 \\
 & MOSS-Audio-8B-Thinking~\cite{mossaudio} & \underline{29.6} & \underline{38.6} & \underline{25.5} & \underline{18.2} & \underline{31.4} & \textbf{23.6} & 0.1 \\
 & Step-Audio-2-mini-Base~\cite{stepaudio2} & 2.0 & 1.9 & 0.7 & 0.1 & 2.3 & 1.1 & 55.5 \\
 & Step-Audio-2-mini~\cite{stepaudio2} & 6.3 & 7.5 & 2.1 & 0.6 & 6.7 & 4.9 & 24.1 \\
 & Step-Audio-2-mini-Think~\cite{stepaudio2} & 2.7 & 2.1 & 0.9 & 0.4 & 3.0 & 1.7 & 27.6 \\
 & Step-Audio-R1~\cite{stepaudior1} & 11.0 & 15.2 & 6.6 & 2.0 & 11.4 & 9.2 & 0.1 \\
 & LLaMA-Omni2-14B~\cite{llamaomni2} & 1.6 & 1.9 & 0.6 & 0.2 & 1.9 & 0.7 & 60.0 \\
\cdashline{1-9}
\multirow{2}{*}{Omni} & LLaMA-3.1-8B-Omni~\cite{llamaomni} & 0.6 & 0.6 & 0.2 & 0.2 & 0.6 & 0.4 & 69.2 \\
 & Gemini-3.1-Pro~\cite{gemini31} & 26.4 & 35.6 & 24.0 & 13.7 & 27.9 & 20.9 & 0.1 \\
\cdashline{1-9}
\multirow{2}{*}{Expert} & LAT-Audio~\cite{lataudio} & 19.7 & 27.0 & 16.2 & 8.3 & 20.7 & 16.2 & 0.2 \\
 & TimeAudio~\cite{timeaudio} & 12.4 & 13.7 & 6.9 & 3.1 & 12.5 & 12.2 & 2.1 \\
\bottomrule
\end{tabular}
\end{adjustbox}
\vspace{-0.5cm}
\end{table*}

\vspace{-0.2cm}
\subsection{Evaluation Protocol}
\vspace{-0.1cm}
\label{ssec:metrics}
\textbf{Answer parsing.} We evaluate a free-form-output track. Timestamps are recovered by a deterministic rule-based cascade that matches interval expressions, start/end patterns, and timestamp arrays, normalizing formats such as \texttt{hh:mm:ss}, \texttt{mm:ss}, millisecond fields, and spelled-out time words into seconds. Model-specific deterministic normalization handles recurring surface conventions, but it does not infer a missing interval from a semantic description. We do not use an LLM parser or a semantic rescue step. If the cascade recovers no valid interval, the response is marked as a parsing failure and normalized to the empty-prediction sentinel \texttt{0s\textasciitilde0s}. The parsing-failure rate is $N_{\mathrm{fail}}/N$. Fail includes any response without a recoverable timestamp, such as a refusal or a no-event statement, and should not be read as a parser malfunction alone. Every failed sample remains in the evaluation: it contributes IoU 0, a miss at every Recall threshold, gIoU $-1$, and zero predicted intervals for the count diagnostics. Consequently, the reported scores measure temporal grounding together with compliance with the timestamp-output convention; they are not format-independent estimates of latent grounding ability.

\textbf{Metrics.} Both prediction and ground truth are treated as interval sets. Overlapping intervals within each side are merged, and we compute the \emph{interval-union IoU}
\begin{equation}
\mathrm{IoU}(\hat{Y},Y)=\frac{|\hat{Y} \cap Y|}{|\hat{Y} \cup Y|},
\label{eq:iou}
\end{equation}
where intersection and union are measured in seconds of the merged interval sets. This formulation handles one-to-many answers natively: missing an occurrence and hallucinating an extra one both lower the score. We report the \textbf{mean IoU (mIoU)} and \textbf{Recall@$\tau$} for $\tau\in\{0.3,0.5,0.7\}$, i.e., the percentage of queries with IoU$\geq\tau$. Throughout the paper, all scores are expressed on a 0--100 scale. As a complementary diagnostic for one-to-many queries, we report \emph{count accuracy} (CntAcc), the percentage of queries for which the number of predicted intervals equals the number of ground-truth intervals. Let $\mathcal{M}(\cdot)$ denote the merging of overlapping intervals, and let $\hat{Y}_i$ and $Y_i$ be the predicted and ground-truth interval sets of the $i$-th query in an evaluation group of $N$ queries; then
\begin{equation}
\mathrm{CntAcc}=\frac{100}{N}\sum_{i=1}^{N}\mathds{1}\!\left[\,\big|\mathcal{M}(\hat{Y}_i)\big|=\big|Y_i\big|\,\right],
\label{eq:cntacc}
\end{equation}
where $|\cdot|$ counts intervals and $\mathds{1}[\cdot]$ is the indicator function. Replacing the equality in Eq.~\eqref{eq:cntacc} with $<$ or $>$ yields the under- and over-reporting rates, which by construction satisfy $\mathrm{CntAcc}+\mathrm{Under}+\mathrm{Over}=100$. A response with no recoverable timestamp contributes zero predicted intervals and is therefore counted as under-reporting. CntAcc deliberately ignores interval positions: combined with the union IoU of Eq.~\eqref{eq:iou}, it separates ``finding all occurrences'' from ``placing boundaries precisely.''

\textbf{Generalized IoU.} The union IoU is zero whenever prediction and ground truth do not overlap, and therefore cannot distinguish a near miss from a widely displaced answer. Following the generalized IoU idea from object detection, we additionally report the quantity in Eq.~\eqref{eq:giou}:
\begin{equation}
\mathrm{gIoU}(\hat{Y},Y)=\mathrm{IoU}(\hat{Y},Y)-\frac{|C|-|\hat{Y}\cup Y|}{|C|},
\label{eq:giou}
\end{equation}
where $C$ is the smallest single interval enclosing both $\hat{Y}$ and $Y$, and $|\cdot|$ measures duration in seconds. The penalty term is the fraction of the enclosure not covered by either interval set, so gIoU ranges over $[-1,1]$: it coincides with the IoU when the prediction is close to the target and decreases toward $-1$ as the prediction drifts away along the timeline; a sample with no recoverable prediction is assigned $-1$.

\textbf{Midpoint absolute error.} As a threshold-free measure of localization bias, we compute the midpoint absolute error (\textbf{MAE}). For a predicted interval with midpoint $\hat{c}=(\hat{s}+\hat{e})/2$ and a ground-truth interval with midpoint $c=(s+e)/2$, the error is $|\hat{c}-c|$ seconds. On one-to-many queries, each of the $M$ ground-truth intervals is matched to the nearest predicted midpoint among the $N$ predictions, and the sample-level MAE is the mean of the $M$ resulting errors. Unlike the other metrics, MAE is computed only over samples with at least one recoverable prediction. We therefore report \textbf{MAE coverage} (Cvg), the number of samples entering the MAE mean divided by the group size. Under the empty-prediction convention above, $\mathrm{Cvg}=1-\mathrm{Fail}$. MAE measures calibration conditional on producing a parseable timestamp and must be read jointly with Cvg rather than as a full-test-set average.

\begin{table*}[t!]
\centering
\caption{Per-subset mIoU. \emph{Event}=acoustic event, \emph{Desc.}=acoustic description, \emph{Rough}=rough semantic, \emph{L-Sem.}/\emph{L-Desc.}=long-audio semantic / acoustic description. All values are percentages. Best per column in \textbf{bold}, second best \underline{underlined}.}
\vspace{-0.2cm}
\label{tab:subset}
\setlength{\tabcolsep}{8pt}
\renewcommand{\arraystretch}{0.86}
\begin{adjustbox}{max width=\textwidth}
\begin{tabular}{c|lcccccccc}
\toprule
Type & Model & Word & Event & Desc. & Semantic & Rough & Emotion & L-Sem. & L-Desc. \\
\midrule
\multirow{17}{*}{GALLMs.} & Audio Flamingo 2 & 1.6 & 0.6 & 0.1 & 1.2 & 0.0 & 0.0 & 0.0 & 0.0 \\
 & AF2 Sound-CoT & 2.8 & 6.3 & 6.7 & 1.8 & 0.8 & 0.3 & 0.1 & 0.3 \\
 & Audio Flamingo 3 & 8.4 & 18.7 & 15.9 & 7.0 & 7.8 & 2.5 & 0.2 & 0.7 \\
 & FireRedAudio & \textbf{34.5} & 31.0 & 36.8 & \textbf{38.4} & \textbf{40.6} & \textbf{25.2} & \textbf{20.6} & \textbf{21.7} \\
 & GLM-4-Voice-9B & 0.8 & 5.2 & 10.5 & 4.0 & 2.6 & 2.7 & 0.0 & 0.0 \\
 & Kimi-Audio-7B & 1.7 & 1.5 & 2.0 & 2.7 & 1.7 & 0.4 & 0.7 & 0.2 \\
 & MiDashengLM-7B & 0.4 & 10.9 & 11.2 & 5.6 & 3.4 & 1.1 & 0.1 & 0.0 \\
 & MiMo-Audio-7B-Instruct & 3.7 & 22.2 & 23.7 & 16.0 & 13.5 & 7.3 & 4.2 & 2.8 \\
 & MOSS-Audio-4B-Instruct & 21.0 & 30.2 & 27.1 & 10.6 & 6.9 & 3.1 & 0.4 & 0.2 \\
 & MOSS-Audio-4B-Thinking & \underline{24.3} & \textbf{51.9} & \textbf{50.6} & 33.9 & 25.7 & 15.7 & 3.2 & 1.2 \\
 & MOSS-Audio-8B-Instruct & 18.0 & 26.0 & 19.1 & 11.5 & 9.1 & 4.3 & 1.6 & 0.6 \\
 & MOSS-Audio-8B-Thinking & 22.9 & \underline{51.9} & \underline{50.0} & \underline{37.4} & \underline{36.5} & \underline{18.2} & 4.8 & 4.1 \\
 & Step-Audio-2-mini-Base & 7.5 & 1.0 & 2.1 & 2.2 & 1.0 & 0.7 & 0.0 & 0.0 \\
 & Step-Audio-2-mini & 12.7 & 7.8 & 9.1 & 7.8 & 6.4 & 2.5 & 0.0 & 0.0 \\
 & Step-Audio-2-mini-Think & 8.5 & 2.6 & 2.6 & 0.8 & 2.9 & 1.5 & 0.0 & 0.0 \\
 & Step-Audio-R1 & 9.5 & 18.7 & 20.7 & 13.6 & 11.8 & 4.0 & 1.3 & 1.6 \\
 & LLaMA-Omni2-14B & 4.4 & 0.0 & 0.1 & 3.2 & 4.4 & 1.7 & 0.5 & 0.0 \\
\cdashline{1-10}
\multirow{2}{*}{Omni} & LLaMA-3.1-8B-Omni & 0.3 & 0.0 & 0.1 & 2.2 & 2.4 & 0.6 & 0.4 & 0.0 \\
 & Gemini-3.1-Pro & 14.3 & 46.5 & 48.1 & 31.6 & 26.8 & 11.4 & \underline{12.9} & \underline{7.2} \\
\cdashline{1-10}
\multirow{2}{*}{Expert} & LAT-Audio & 3.2 & 31.9 & 31.3 & 35.6 & 29.6 & 13.4 & 8.6 & 5.5 \\
 & TimeAudio & 7.0 & 23.5 & 24.3 & 15.7 & 12.4 & 9.0 & 1.6 & 0.6 \\
\bottomrule
\end{tabular}
\end{adjustbox}
\vspace{-0.5cm}
\end{table*}

\vspace{-0.2cm}
\section{Results and Analysis}
\vspace{-0.1cm}
\label{sec:results}
\vspace{-0.2cm}
\subsection{Main Results}
\vspace{-0.1cm}
Table~\ref{tab:main} reports the overall results, and Table~\ref{tab:subset} breaks them down by subset; complete per-subset tables including all recall thresholds, together with supplementary figures, are provided in the Appendix (Figs.~\ref{fig:heatmap}--\ref{fig:countacc}, Tables~\ref{tab:appx1}--\ref{tab:appx4}). Table~\ref{tab:main} supports three key conclusions.

Temporal grounding remains far from solved. FireRedAudio has the highest point estimate with 31.2 mIoU, R@0.5 of 30.8, and R@0.7 of 21.5. It is followed by MOSS-Audio-8B-Thinking and MOSS-Audio-4B-Thinking at 29.6 and 27.8 mIoU, respectively. Nine of the 21 models score below 5 mIoU, and even the highest-scoring system localizes only about one fifth of queries at IoU$\geq$0.7. Given that the median target covers only 7\% of its recording, the strict-overlap results show that high-precision localization remains difficult.

The two approximately 17-minute subsets remain difficult. FireRedAudio leads with 20.6 mIoU on long semantic queries and 21.7 on long acoustic descriptions, ahead of Gemini-3.1-Pro (12.9/7.2) and LAT-Audio (8.6/5.5). This empirical profile is consistent with FireRedAudio's stated support for timestamped long-form audio understanding~\cite{fireredaudio}. Nevertheless, the best mIoU in these subsets is only 21.7, and FireRedAudio fails to produce a recoverable timestamp on 20.8\% of long acoustic-description queries. Because the long and short subsets also differ in source and query distribution, these results establish performance in the 17--20-minute regime but do not isolate duration as the cause of the gap.

One-to-many queries are harder for the leading systems. One-to-many mIoU falls below one-to-one mIoU for FireRedAudio (34.1 vs.\ 21.0), MOSS-Audio-8B-Thinking (31.4 vs.\ 23.6), and Gemini-3.1-Pro (27.9 vs.\ 20.9). At the stricter threshold, the best R@0.7 across the evaluated models falls from 24.9 on one-to-one queries to 10.1 on one-to-many queries. Sec.~\ref{ssec:o2m} dissects this failure mode.

\vspace{-0.2cm}
\subsection{Subset-Level Variation Across Query Categories}
\vspace{-0.1cm}
The short-audio subsets distinguish fine boundary placement from broader acoustic-event matching. On word queries, general-purpose audio LLMs occupy the leading positions: FireRedAudio reaches 34.5 mIoU, followed by MOSS-Audio-4B-Thinking and MOSS-Audio-8B-Thinking at 24.3 and 22.9, whereas the strongest omni-modal and expert results are 14.3 from Gemini-3.1-Pro and 7.0 from TimeAudio. Even FireRedAudio drops from 47.7 R@0.3 to 25.8 R@0.7, confirming that precise word boundaries remain difficult. On acoustic event and description, the MOSS thinking variants lead at 51.9 and 50.6 mIoU, with Gemini-3.1-Pro close at 46.5/48.1 and the best expert-model scores at 31.9/31.3 from LAT-Audio. Within the MOSS family, thinking improves event/description mIoU by approximately 22--31 points over the corresponding instruct variants, but improves word mIoU by only 3--5 points. The observed thinking--instruct gap is therefore much larger for acoustic-content matching than for short word boundaries.

The semantic and emotion subsets produce a different cross-group profile. FireRedAudio leads semantic, rough-semantic, and emotion grounding at 38.4, 40.6, and 25.2 mIoU, followed by MOSS-Audio-8B-Thinking at 37.4, 36.5, and 18.2. LAT-Audio remains competitive on semantic and rough-semantic queries (35.6/29.6), while Gemini-3.1-Pro reaches 31.6/26.8; neither the expert nor omni-modal model leads these subsets. Emotion has a lower ceiling across all three groups: its best score is 25.2, compared with 38.4 and 40.6 for semantic and rough-semantic grounding. Moreover, rough-semantic queries are easier than semantic queries for FireRedAudio but harder for MOSS-Audio-8B-Thinking, LAT-Audio, and Gemini-3.1-Pro, so the results do not support a single ordering by query abstraction.

Long audio reshuffles the short-audio ranking. FireRedAudio, a general-purpose audio LLM, leads both long subsets at 20.6/21.7 mIoU; Gemini-3.1-Pro is second at 12.9/7.2, and the expert LAT-Audio reaches 8.6/5.5. In contrast, MOSS-Audio-8B-Thinking, which leads or nearly leads the short acoustic subsets, falls to 4.8/4.1 on long audio. Strong short-audio event grounding and an expert-model designation therefore do not by themselves predict long-context performance under this protocol. These group-level patterns remain descriptive: source domain, duration, and one-to-many prevalence vary across subsets, so they should not be interpreted as causal effects of model architecture or model type.

\begin{table*}[t!]
\centering
\caption{Generalized IoU (gIoU, scaled to $[-100,100]$, higher is better), midpoint absolute error (MAE, seconds, lower is better), and MAE coverage (Cvg, percentage), overall and split into one-to-one and one-to-many queries. MAE is computed only over samples with at least one recoverable prediction; Cvg reports the fraction entering that mean. Best and second-best gIoU/MAE values are in \textbf{bold} and \underline{underlined}; Cvg is not ranked.}
\vspace{-0.2cm}
\label{tab:gioumae}
\setlength{\tabcolsep}{8pt}
\renewcommand{\arraystretch}{0.86}
\begin{adjustbox}{max width=\textwidth}
\begin{tabular}{c|lccc|ccc|ccc}
\toprule
& & \multicolumn{3}{c|}{Overall} & \multicolumn{3}{c|}{One-to-one} & \multicolumn{3}{c}{One-to-many} \\
Type & Model & gIoU & MAE & Cvg & gIoU & MAE & Cvg & gIoU & MAE & Cvg \\
\midrule
\multirow{17}{*}{GALLMs.} & Audio Flamingo 2 & -69.2 & 149.6 & 99.9 & -67.7 & 149.8 & 100.0 & -74.5 & 148.9 & 99.7 \\
 & AF2 Sound-CoT & -55.9 & 146.1 & 99.9 & -55.5 & 146.4 & 99.9 & -57.4 & 144.8 & 100.0 \\
 & Audio Flamingo 3 & -38.5 & 144.7 & 99.9 & -37.2 & 146.2 & 99.9 & -43.2 & 139.4 & 100.0 \\
 & FireRedAudio & -0.6 & \textbf{34.3} & 84.9 & \underline{7.9} & \textbf{26.8} & 87.7 & -30.4 & 65.5 & 74.7 \\
 & GLM-4-Voice-9B & -61.4 & 305.6 & 66.5 & -61.6 & 331.5 & 65.4 & -60.5 & 220.3 & 70.0 \\
 & Kimi-Audio-7B & -69.6 & 1101.4 & 73.1 & -67.6 & 863.2 & 75.3 & -76.8 & 2068.3 & 65.4 \\
 & MiDashengLM-7B & -57.6 & 85.6 & 71.3 & -58.4 & 91.0 & 69.9 & -55.0 & 68.2 & 76.0 \\
 & MiMo-Audio-7B-Instruct & -30.7 & 365.6 & 98.1 & -28.2 & 349.7 & 98.4 & -39.4 & 422.7 & 96.9 \\
 & MOSS-Audio-4B-Instruct & -28.5 & 286.2 & 96.5 & -24.8 & 269.6 & 96.6 & -41.7 & 345.1 & 96.1 \\
 & MOSS-Audio-4B-Thinking & -1.5 & 88.2 & 99.1 & 3.2 & 87.0 & 99.0 & -18.0 & 92.6 & 99.2 \\
 & MOSS-Audio-8B-Instruct & -32.8 & 159.1 & 88.8 & -27.8 & 168.5 & 90.6 & -50.3 & 122.6 & 82.4 \\
 & MOSS-Audio-8B-Thinking & \textbf{3.5} & 923.1 & 99.9 & \textbf{8.7} & 943.6 & 100.0 & \textbf{-14.6} & 851.1 & 99.7 \\
 & Step-Audio-2-mini-Base & -72.4 & 42.9 & 44.5 & -69.3 & 43.3 & 46.8 & -83.6 & \underline{40.7} & 36.4 \\
 & Step-Audio-2-mini & -44.1 & \underline{36.4} & 75.9 & -40.4 & \underline{37.7} & 75.6 & -56.8 & \textbf{32.0} & 77.3 \\
 & Step-Audio-2-mini-Think & -59.7 & 230.9 & 72.4 & -56.5 & 223.5 & 72.2 & -71.2 & 256.9 & 73.1 \\
 & Step-Audio-R1 & -30.0 & 317.3 & 99.9 & -27.7 & 359.1 & 99.9 & -38.1 & 169.6 & 99.7 \\
 & LLaMA-Omni2-14B & -78.8 & 118.2 & 40.0 & -75.4 & 121.9 & 45.4 & -90.7 & 89.6 & 20.9 \\
\cdashline{1-11}
\multirow{2}{*}{Omni} & LLaMA-3.1-8B-Omni & -88.8 & 340.8 & 30.8 & -87.6 & 354.5 & 34.7 & -93.1 & 242.3 & 17.1 \\
 & Gemini-3.1-Pro & \underline{0.5} & 61.2 & 99.9 & 5.0 & 57.2 & 99.9 & \underline{-15.4} & 75.4 & 100.0 \\
\cdashline{1-11}
\multirow{2}{*}{Expert} & LAT-Audio & -7.0 & 58.0 & 99.8 & -2.8 & 53.3 & 99.7 & -21.6 & 74.4 & 100.0 \\
 & TimeAudio & -16.0 & 127.9 & 97.9 & -15.8 & 126.8 & 97.7 & -16.7 & 131.8 & 98.4 \\
\bottomrule
\end{tabular}
\end{adjustbox}
\vspace{-0.5cm}
\end{table*}

\vspace{-0.2cm}
\subsection{Beyond Overlap: gIoU and Midpoint Error}
\vspace{-0.1cm}
\label{ssec:gioumae}

Table~\ref{tab:gioumae} shows that gIoU changes the interpretation of the mIoU ranking. FireRedAudio leads in mIoU (31.2) but ranks third in overall gIoU ($-0.6$), behind MOSS-Audio-8B-Thinking (3.5) and Gemini-3.1-Pro (0.5). Its parseable responses average 17.1 gIoU, but the 15.1\% parsing failures, each scored as $-100$, pull the full-set mean below zero. MOSS-Audio-8B-Thinking remains positive with 99.9\% coverage. Moreover, every model has negative one-to-many gIoU, with a best value of only $-14.6$, showing that recurring-target errors extend beyond small boundary offsets to missing or displaced occurrences.

MAE must likewise be read with coverage and overlap. FireRedAudio has the lowest MAE (34.3\,s) at 84.9\% coverage, whereas Step-Audio-2-mini and its Base variant reach similar MAE values (36.4/42.9\,s) but only 6.3/2.0 mIoU and 75.9\%/44.5\% coverage. Low parsed-only MAE therefore does not imply strong full-benchmark grounding. Conversely, MOSS-Audio-8B-Thinking has the best gIoU but a 923.1\,s MAE, with 96.8\% of its summed midpoint error coming from the two long-audio subsets. This long-audio error tail is visible in the unbounded, seconds-based MAE but is compressed by bounded gIoU.

These metrics capture different failure modes. FireRedAudio has the highest point-estimate overlap and lowest parsed-only MAE but a 15.1\% output-failure rate; MOSS-Audio-8B-Thinking has the highest gIoU and almost complete coverage but large long-audio errors; and no model achieves positive one-to-many gIoU. We therefore use mIoU and Recall as the primary full-benchmark measures, gIoU to diagnose displacement, and MAE only together with Cvg.

\vspace{-0.2cm}
\subsection{One-to-One versus One-to-Many Grounding}
\vspace{-0.1cm}
\label{ssec:o2m}

Fig.~\ref{fig:single_multi} (Appendix) plots one-to-one against one-to-many mIoU for all 21 models, and Table~\ref{tab:count} reports the interval-count diagnostics. Together they identify \emph{under-reporting} as the dominant failure mode on one-to-many queries.

\begin{table}[t!]
\centering
\caption{Interval-count diagnostics. Left block: one-to-many queries (n=387, 4.02 ground-truth intervals on average): count accuracy, under- and over-reporting rates, and the average number of predicted intervals. Right block: one-to-one queries (n=1{,}363): count accuracy and over-reporting rate. All rates are percentages.}
\vspace{-0.2cm}
\label{tab:count}
\setlength{\tabcolsep}{2.6pt}
\resizebox{\columnwidth}{!}{%
\begin{tabular}{c|lcccc|cc}
\toprule
& & \multicolumn{4}{c|}{One-to-many} & \multicolumn{2}{c}{One-to-one} \\
Type & Model & CntAcc & Under & Over & \#Pred & CntAcc & Over \\
\midrule
\multirow{17}{*}{GAU} & Audio Flamingo 2 & 0.3 & 98.4 & 1.3 & 1.18 & 99.3 & 0.7 \\
 & AF2 Sound-CoT & 0.3 & 99.5 & 0.3 & 1.07 & 99.8 & 0.1 \\
 & Audio Flamingo 3 & 0.3 & 99.7 & 0.0 & 1.00 & 99.7 & 0.2 \\
 & FireRedAudio & 4.7 & 91.0 & 4.4 & 1.19 & 84.2 & 3.5 \\
 & GLM-4-Voice-9B & 0.3 & 99.2 & 0.5 & 0.73 & 63.5 & 1.9 \\
 & Kimi-Audio-7B & 1.0 & 97.2 & 1.8 & 0.95 & 73.6 & 1.8 \\
 & MiDashengLM-7B & 1.3 & 98.4 & 0.3 & 0.79 & 69.3 & 0.6 \\
 & MiMo-Audio-7B-Instruct & 1.0 & 99.0 & 0.0 & 0.98 & 97.7 & 0.7 \\
 & MOSS-Audio-4B-Instruct & 0.3 & 99.7 & 0.0 & 0.97 & 96.6 & 0.1 \\
 & MOSS-Audio-4B-Thinking & 3.6 & 95.3 & 1.0 & 1.22 & 97.4 & 1.7 \\
 & MOSS-Audio-8B-Instruct & 0.3 & 99.7 & 0.0 & 0.82 & 90.5 & 0.1 \\
 & MOSS-Audio-8B-Thinking & 1.3 & 97.9 & 0.8 & 1.11 & 98.5 & 1.5 \\
 & Step-Audio-2-mini-Base & 0.0 & 100.0 & 0.0 & 0.36 & 46.8 & 0.0 \\
 & Step-Audio-2-mini & 0.0 & 100.0 & 0.0 & 0.78 & 75.6 & 0.0 \\
 & Step-Audio-2-mini-Think & 1.3 & 98.4 & 0.3 & 0.76 & 69.1 & 3.1 \\
 & Step-Audio-R1 & 13.2 & 79.3 & 7.5 & 1.65 & 66.8 & 33.2 \\
 & LLaMA-Omni2-14B & 0.8 & 99.2 & 0.0 & 0.21 & 44.6 & 0.8 \\
\cdashline{1-8}
\multirow{2}{*}{Omni} & LLaMA-3.1-8B-Omni & 0.8 & 99.2 & 0.0 & 0.18 & 34.6 & 0.1 \\
 & Gemini-3.1-Pro & 7.0 & 82.1 & 10.9 & 3.67 & 92.7 & 7.1 \\
\cdashline{1-8}
\multirow{2}{*}{Expert} & LAT-Audio & 0.3 & 99.7 & 0.0 & 1.00 & 99.7 & 0.0 \\
 & TimeAudio & 3.6 & 85.8 & 10.6 & 2.38 & 82.3 & 15.4 \\
\bottomrule
\end{tabular}}
\vspace{-0.5cm}
\end{table}

Models usually find one occurrence and stop. One-to-many queries contain 4.02 ground-truth intervals on average, whereas average predictions range from 0.18 to 3.67 intervals and under-reporting rates range from 79.3\% to 100\% (Fig.~\ref{fig:countacc}, Appendix). Most systems emit roughly one interval or fewer. Gemini-3.1-Pro is the exception in output volume at 3.67 predictions, but its count accuracy is still only 7.0\% because the predicted number is seldom exactly correct. FireRedAudio averages 1.19 predictions and under-reports 91.0\% of one-to-many queries despite leading the overall localization metrics. The task definition requires all matching intervals, but the item question does not disclose the target count or whether the example is one-to-many; count accuracy therefore measures recurrence discovery and enumeration without an oracle count.

Occurrence counting remains weak under this protocol. The highest one-to-many count accuracy is 13.2\% (Step-Audio-R1), followed by Gemini-3.1-Pro at 7.0\%, FireRedAudio at 4.7\%, and MOSS-Audio-4B-Thinking and TimeAudio at 3.6\%. Step-Audio-R1's lead coincides with a 33.2\% over-reporting rate on one-to-one queries, indicating that its tendency to emit several intervals sometimes matches a multi-interval ground truth by chance. No model exceeds 13.2\% count accuracy, so the results provide little evidence of reliable occurrence counting.

Apparent one-to-many advantages can be artifacts. A few weak models (MiDashengLM-7B, GLM-4-Voice-9B) score marginally higher on one-to-many than on one-to-one queries. Inspection shows that their predictions are systematically long intervals; against a multi-interval ground truth whose union is larger, a long prediction accrues intersection by accident. Such cases underline why mIoU alone is insufficient and why the count diagnostics of Table~\ref{tab:count} are needed to separate recall of occurrences from boundary precision.

\vspace{-0.2cm}
\subsection{Model-Series Analysis}
\vspace{-0.1cm}
Reasoning gains are family-dependent. Within MOSS-Audio, the thinking variants improve substantially over their instruct counterparts (8B: 29.6 vs.\ 12.8; 4B: 27.8 vs.\ 14.6), and the 4B--8B gap is smaller than the thinking--instruct gap. The Step-Audio family does not reproduce this pattern uniformly: Step-Audio-2-mini-Think scores 2.7, below the non-thinking mini at 6.3, although Step-Audio-R1 reaches 11.0. Sound-CoT improves Audio Flamingo 2 from 0.5 to 2.9 but remains weak in absolute terms. The results therefore do not support a family-independent benefit from explicit reasoning.

FireRedAudio has a distinct benchmark profile. It has the highest overall point estimate (31.2) and leads both long subsets (20.6/21.7), consistent with its stated long-audio temporal-grounding capability~\cite{fireredaudio}. Its 15.1\% overall Fail rate and 91.0\% one-to-many under-reporting rate show that this profile does not produce uniformly recoverable answers or complete occurrence enumeration.

Expert models remain competitive but do not lead the two long subsets. LAT-Audio reaches 19.7 overall and 8.6/5.5 on the two long subsets; its whole-second outputs also limit word-level mIoU to 3.2. TimeAudio reaches 12.4 overall and drops to 3.1 at R@0.7. These results show that specialization alone does not guarantee either fine boundary precision or the strongest long-context performance.

The commercial Gemini-3.1-Pro baseline is competitive but exhibits a different trade-off from the leading audio models. It reaches 26.4 mIoU overall and ranks second on both long subsets at 12.9/7.2. On one-to-many queries it emits 3.67 intervals on average, producing the second-best count accuracy of 7.0\%, while still under-reporting 82.1\% of the cases. Higher output volume therefore does not by itself produce reliable occurrence enumeration.

The two open-source LLaMA-Omni systems perform poorly under this protocol. LLaMA-3.1-8B-Omni scores 0.6 mIoU and fails parsing on 69.2\% of queries; LLaMA-Omni2-14B scores 1.6 with 60.0\% Fail. These results show that the evaluated systems rarely produce correct explicit temporal answers in the free-form setting, but their grounding and format-compliance errors cannot be fully separated.

Model selection is metric-dependent rather than governed by a single leaderboard. FireRedAudio combines the highest overall mIoU (31.2) with the smallest parsed-only MAE (34.3\,s), but its 84.9\% MAE coverage leaves 15.1\% of the benchmark without a recoverable timestamp. MOSS-Audio-8B-Thinking has the highest overall gIoU (3.5) and 99.9\% coverage, while its parsed-only MAE is 923.1\,s. Step-Audio-R1 has the highest one-to-many count accuracy (13.2\%) but reaches only 11.0 mIoU overall. These profiles isolate different requirements: output-format reliability, boundary placement, and occurrence enumeration. A complete comparison should therefore report full-coverage overlap metrics, parsed-only MAE with Cvg, and count diagnostics together rather than treating any one rank as sufficient.

\vspace{-0.2cm}
\subsection{Scope and Limitations}
\vspace{-0.1cm}
\label{ssec:scope}
\textbf{Query-level sample size.} 
TAG-Bench contains 1{,}750 distinct query--recording pairs, and each recording contributes exactly one query. Subset sizes range from 129 emotion queries to 301 acoustic-event queries; the two long-audio subsets contain 180 semantic and 260 acoustic-description queries. The evaluation therefore summarizes one grounding decision per recording but cannot measure whether a model behaves consistently across different queries about the same audio. Adding multiple queries to each long recording would enable within-recording comparisons without requiring a proportional increase in audio duration.

\textbf{Aggregation and weighting.} Overall metrics are macro-averaged over query--recording pairs, so each query contributes equally regardless of recording duration or the number of ground-truth intervals. The 440 long-audio queries therefore contribute 25.1\% of the overall score even though their recordings account for 127.6 of the benchmark's 149.5 hours. This avoids weighting a query by audio duration, but it also means that the overall ranking reflects the benchmark's sample mixture. Per-subset and one-to-many results should therefore accompany the overall score when comparing models for a specific use case.

\textbf{Source and duration confounding.} The eight subsets differ in source domain, query type, duration, and recurrence rate. We therefore do not interpret their ordering as a controlled abstraction ladder or the short--long gap as an isolated context-length effect. In addition, the longest recording is 20 minutes. Evaluation on matched recordings with multiple query categories, together with 30-minute and hour-scale material, is needed before extrapolating the findings to longer contexts.

\textbf{Free-form output confounding.} The deterministic parser makes scoring reproducible and exposes Fail rather than silently dropping responses without timestamps, but the resulting metric combines temporal grounding with instruction following and timestamp formatting. The uniform item template does not provide an oracle occurrence count; an explicit reminder to enumerate all intervals is therefore an untested prompt intervention. The parsed-only arithmetic check above preserves the leading order, but it is selection-biased and does not establish that rankings would remain fixed under constrained JSON decoding or model-specific prompt optimization. A paired free-form and constrained-output track would directly measure that sensitivity.
\vspace{-0.2cm}
\section{Conclusion}
\vspace{-0.1cm}
\label{sec:conclusion}
We have introduced TAG-Bench, a human-verified benchmark for temporal audio grounding comprising 1{,}750 query--recording pairs over 149.5 hours of audio, eight source-dependent diagnostic subsets, and a union-IoU protocol with count diagnostics for one-to-many targets. Across 21 systems, FireRedAudio achieves the highest overall mIoU at 31.2 and reaches 20.6/21.7 mIoU on the two approximately 17-minute subsets, but its R@0.7 is only 21.5 and it under-reports 91.0\% of one-to-many queries. No model exceeds 13.2\% one-to-many count accuracy, showing that complete occurrence recovery remains difficult. Parsing failures vary widely across systems; they count as misses for mIoU, Recall, gIoU, and count metrics but are excluded from MAE, making MAE coverage necessary for interpretation. We will release the TAG-Bench data and evaluation code to support future research.

% Keep the bibliography on a separate page from the main text.
\clearpage
% References
% -------------------------------------------------------------------------
\bibliographystyle{IEEEbib}
\bibliography{strings,refs}

% -------------------------------------------------------------------------
% ----------------------------    Appendix     ----------------------------
% -------------------------------------------------------------------------
\clearpage
\appendix
\raggedbottom
% Appendix float pages are top-aligned; remaining space is left at the bottom
% instead of being split above and below the tables.
\setcounter{dbltopnumber}{2}
\setcounter{totalnumber}{4}
\renewcommand{\dbltopfraction}{0.98}
\renewcommand{\dblfloatpagefraction}{0.45}
\renewcommand{\topfraction}{0.95}
\renewcommand{\textfraction}{0.05}
\setlength{\dblfloatsep}{6pt plus 1pt minus 1pt}
\setlength{\dbltextfloatsep}{8pt plus 2pt minus 2pt}
\makeatletter
\setlength{\@dblfptop}{0pt}
\setlength{\@dblfpsep}{10pt plus 2pt minus 1pt}
\setlength{\@dblfpbot}{0pt plus 1fil}
\makeatother
% 在导言区加入

\begin{strip}
  \centering
  \vspace{-1.2cm}
  \section*{APPENDICES}
  \vspace{-0.3cm}
\end{strip}
\setlength{\stripsep}{5pt plus 2pt minus 2pt}

% 之后自动回到双栏

\section{TAG-Bench Detail Visualizations}
\begin{figure}[ht!]
  \centering
  \includegraphics[width=\columnwidth]{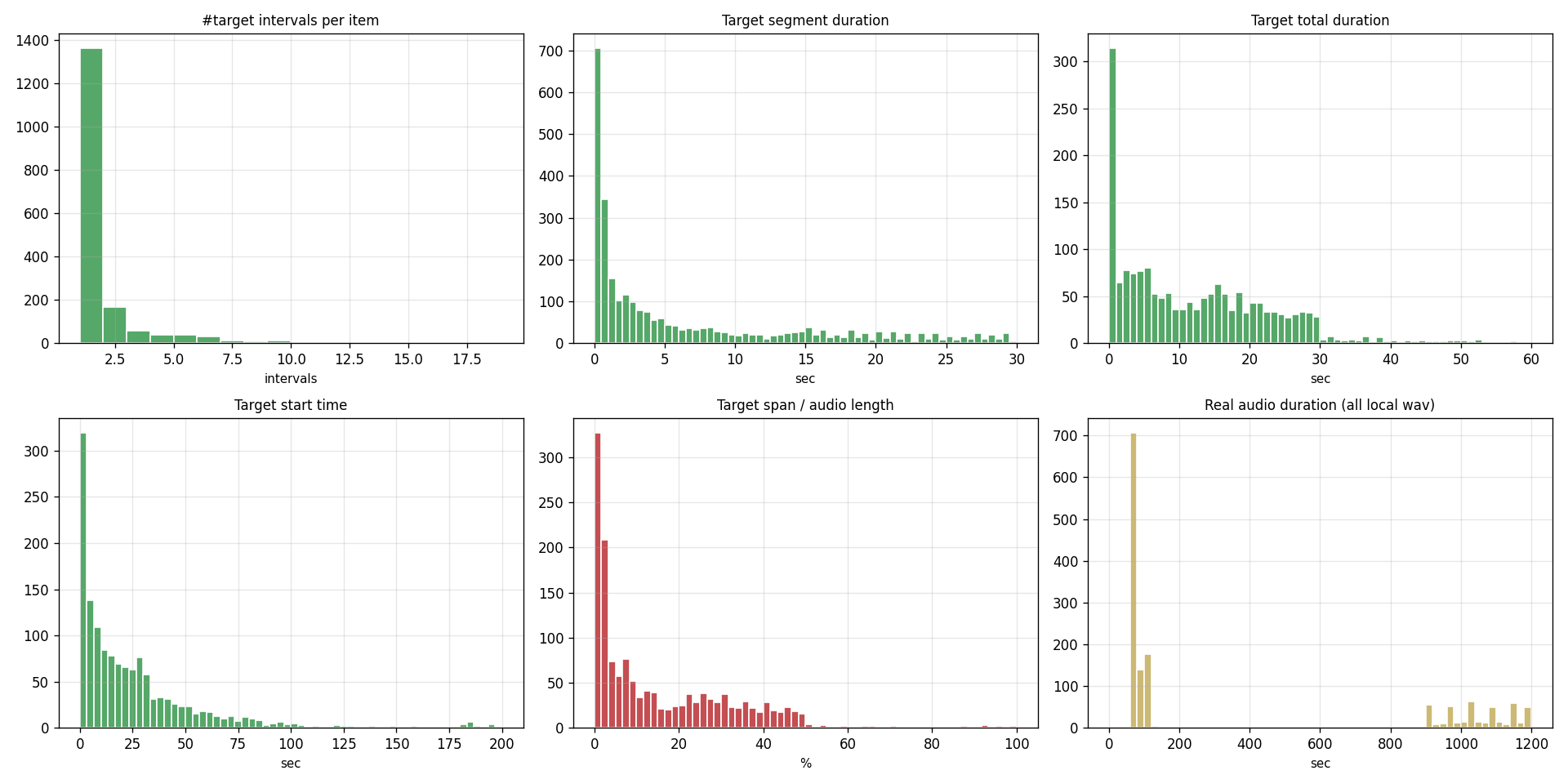}
  \caption{Ground-truth statistics of TAG-Bench: number of target intervals per query, target segment and total duration, target onset, target-to-audio coverage ratio, and audio duration. Targets are short (median 2.7\,s) and sparse (median 7\% coverage). The duration panel uses the 1{,}471 recordings available to the local probe (all non-word items) and displays the full range up to 1{,}200\,s; the other panels include all 1{,}750 queries.}
  \label{fig:stats}
  \vspace{-0.5cm}
\end{figure}

\textbf{Fig.~\ref{fig:stats} (needle-in-a-haystack targets).} As shown in Fig.~\ref{fig:stats}, the median target segment lasts only 2.7\,s, and the median target-to-audio coverage ratio is 7\%: the answer typically occupies a small fraction of the recording. Target onsets range from the first seconds up to 1{,}190\,s, so models cannot succeed by defaulting to the beginning of the audio. The audio-duration distribution is bimodal, separating the short-to-medium subsets from the 17--20-minute subsets.

\FloatBarrier

\section{Supplementary visualizations of test results}
\label{sec:appfigs}
\begin{figure}[t!]
  \centering
  \includegraphics[width=\columnwidth]{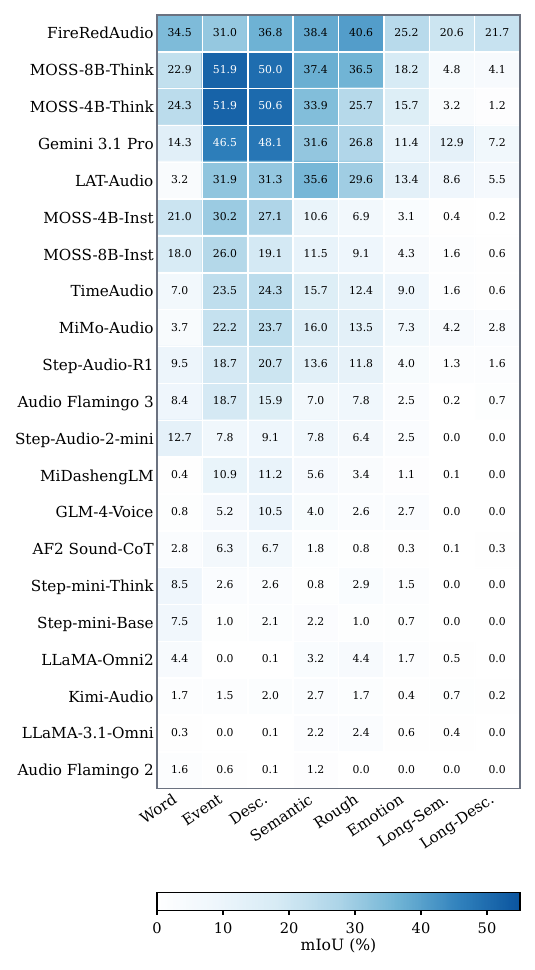}
  \caption{FireRedAudio leads four short-audio subsets and both long-audio subsets, while the MOSS thinking variants lead the two short acoustic subsets. The heatmap reports mIoU across 21 models and eight subsets; darker blue denotes higher mIoU.}
  \label{fig:heatmap}
\end{figure}

\begin{figure}[t!]
  \centering
  \includegraphics[width=\columnwidth]{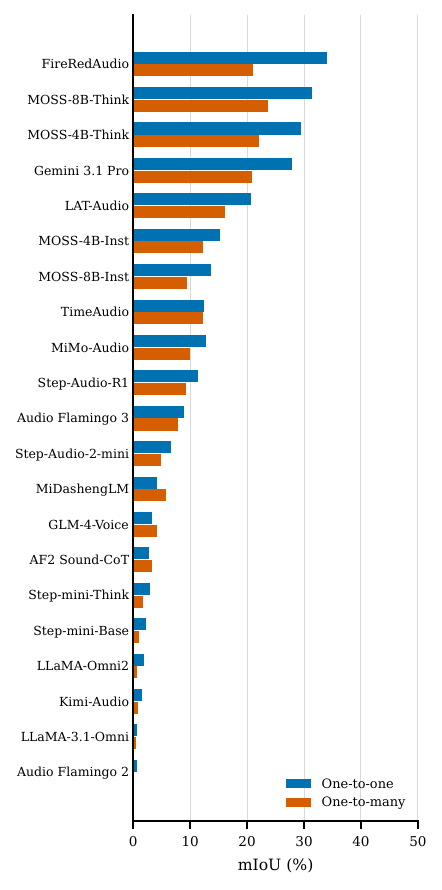}
  \caption{One-to-one vs.\ one-to-many mIoU across the 21 evaluated models. Every competitive model degrades on queries with multiple ground-truth intervals.}
  \label{fig:single_multi}
\end{figure}

\begin{figure}[t!]
  \centering
  \includegraphics[width=\columnwidth]{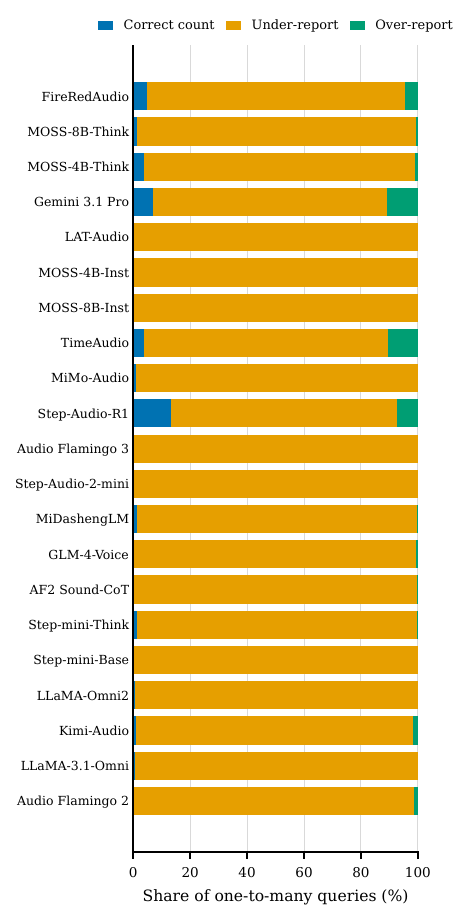}
  \caption{Count accuracy, under-reporting, and over-reporting rates on one-to-many queries. Under-reporting dominates for every evaluated model.}
  \label{fig:countacc}
\end{figure}

Figs.~\ref{fig:heatmap}--\ref{fig:countacc} provide the model--subset mIoU heatmap, the one-to-one versus one-to-many comparison, and the interval-count diagnostics referenced in Sec.~\ref{sec:results}.

\textbf{Fig.~\ref{fig:heatmap} (model--subset heatmap).} The heatmap shows distinct frontier profiles. FireRedAudio leads word, semantic, rough-semantic, and emotion queries and forms the only row above 20 mIoU in both long columns. The MOSS thinking models lead short acoustic event and description queries at approximately 50--52 mIoU but fall below 5 on long audio, while Gemini-3.1-Pro ranks second on both long subsets. The figure therefore separates short-audio acoustic performance from long-form temporal specialization.

\textbf{Fig.~\ref{fig:single_multi} (one-to-one vs.\ one-to-many).} The gap persists among the leading systems: FireRedAudio drops from 34.1 to 21.0, MOSS-Audio-8B-Thinking from 31.4 to 23.6, and Gemini-3.1-Pro from 27.9 to 20.9. The only inversions occur among weak systems, including MiDashengLM-7B (4.2 vs.\ 5.8) and GLM-4-Voice-9B (3.2 vs.\ 4.1); as discussed in Sec.~\ref{ssec:o2m}, their long predictions can intersect a larger multi-interval union without correctly enumerating its occurrences.

\textbf{Fig.~\ref{fig:countacc} (interval-count diagnostics).} Under-reporting dominates all 21 rows (79.3--100\%). Step-Audio-R1 has the largest count-accuracy segment at 13.2\%, followed by Gemini-3.1-Pro (7.0\%) and FireRedAudio (4.7\%). Gemini predicts 3.67 intervals per one-to-many query, close to the 4.02 ground-truth average, but its simultaneous 82.1\% under-reporting and 10.9\% over-reporting show that output volume alone does not recover the correct count. FireRedAudio remains predominantly a single-interval responder despite leading the overall localization scores.

\FloatBarrier

%-------------------------------------------------------------
%--------------------  Appendix B  ---------------------------
%-------------------------------------------------------------
\section{Complete Per-Subset Results}
\label{sec:appendix}
Tables~\ref{tab:appx1}--\ref{tab:appx4} use a common two-panel layout to report every model on all eight subsets. Each panel contains mIoU, Recall at three thresholds, gIoU, parsed-only midpoint MAE, count accuracy (CntAcc), and parsing-failure rate (Fail); model ordering and grouping follow Table~\ref{tab:main}. Except for MAE, values are percentages, and gIoU is scaled to $[-100,100]$. A failed parse is scored as an empty prediction by every full-coverage metric but is excluded from MAE, making the corresponding MAE coverage exactly $100-\mathrm{Fail}$. Best and second-best results are bold and underlined, respectively; Fail is reported diagnostically rather than ranked.
\captionsetup[table]{font=small,skip=3pt}

% Keep two consecutive result tables in each full-width block. This preserves
% table order while preventing a short trailing text-only appendix page.

\begin{strip}
\centering
\vspace{+0.4cm}
\captionof{table}{Complete per-model results for the word and acoustic-event subsets.}
\label{tab:appx1}
\vspace{-0.06cm}
\setlength{\tabcolsep}{2.2pt}
\renewcommand{\arraystretch}{1.2}
\footnotesize
\begin{minipage}[t]{0.49\textwidth}
\vspace{0pt}\centering
\textbf{\strut (a) Word}\par\vspace{1pt}
\resizebox{\linewidth}{!}{%
\begin{tabular}{l*{8}{N}}
\toprule
Model & mIoU & R@.3 & R@.5 & R@.7 & gIoU & MAE & CntAcc & Fail \\
\midrule
Audio Flamingo 2 & 1.6 & 2.9 & 1.1 & 0.0 & -68.0 & 3.0 & \textbf{98.6} & 0.0 \\
AF2 Sound-CoT & 2.8 & 3.9 & 1.4 & 0.7 & -58.5 & 2.7 & \textbf{98.6} & 0.0 \\
Audio Flamingo 3 & 8.4 & 12.2 & 5.7 & 1.4 & -35.3 & 5.0 & \textbf{98.6} & 0.0 \\
FireRedAudio & \textbf{34.5} & \textbf{47.7} & \textbf{42.7} & \textbf{25.8} & 9.5 & 1.0 & 97.1 & 0.7 \\
GLM-4-Voice-9B & 0.8 & 0.0 & 0.0 & 0.0 & -63.4 & 142.9 & 92.8 & 5.4 \\
Kimi-Audio-7B & 1.7 & 1.4 & 0.4 & 0.0 & -62.2 & 2.3 & 91.4 & 6.8 \\
MiDashengLM-7B & 0.4 & 0.7 & 0.0 & 0.0 & -84.1 & 7.1 & 66.7 & 33.0 \\
MiMo-Audio-7B-Instruct & 3.7 & 3.9 & 1.8 & 0.0 & -46.7 & 32.7 & 97.5 & 0.0 \\
MOSS-Audio-4B-Instruct & 21.0 & 33.0 & \underline{15.8} & 4.3 & 3.6 & 0.7 & \textbf{98.6} & 0.0 \\
MOSS-Audio-4B-Thinking & \underline{24.3} & \underline{36.6} & 8.2 & 1.4 & \underline{16.8} & 0.6 & \underline{98.2} & 0.0 \\
MOSS-Audio-8B-Instruct & 18.0 & 27.6 & 15.1 & \underline{5.4} & -1.1 & 0.8 & \textbf{98.6} & 0.0 \\
MOSS-Audio-8B-Thinking & 22.9 & 31.9 & 5.0 & 0.4 & \textbf{20.3} & \textbf{0.4} & \textbf{98.6} & 0.0 \\
Step-Audio-2-mini-Base & 7.5 & 7.5 & 1.8 & 0.4 & -29.7 & 1.6 & 94.6 & 3.9 \\
Step-Audio-2-mini & 12.7 & 15.1 & 3.6 & 1.4 & -12.0 & 1.6 & \textbf{98.6} & 0.0 \\
Step-Audio-2-mini-Think & 8.5 & 5.4 & 2.9 & 1.4 & -11.4 & 4.8 & 92.5 & 4.3 \\
Step-Audio-R1 & 9.5 & 13.6 & 5.7 & 1.8 & -23.7 & 2.3 & 59.1 & 0.0 \\
LLaMA-Omni2-14B & 4.4 & 5.4 & 1.1 & 0.4 & -42.3 & 11.4 & 97.1 & 1.8 \\
\hdashline
LLaMA-3.1-8B-Omni & 0.3 & 0.4 & 0.0 & 0.0 & -91.9 & 21.1 & 39.8 & 60.2 \\
Gemini-3.1-Pro & 14.3 & 21.5 & 10.4 & 1.4 & -4.3 & \underline{0.6} & \underline{98.2} & 0.0 \\
\hdashline
LAT-Audio & 3.2 & 0.7 & 0.4 & 0.0 & -26.6 & 10.0 & \textbf{98.6} & 0.0 \\
TimeAudio & 7.0 & 6.5 & 1.1 & 1.1 & -21.6 & 1.8 & 92.8 & 0.0 \\
\bottomrule
\end{tabular}}

\end{minipage}\hfill
\begin{minipage}[t]{0.49\textwidth}
\vspace{0pt}\centering
\textbf{\strut (b) Acoustic event}\par\vspace{1pt}
\resizebox{\linewidth}{!}{%
\begin{tabular}{l*{8}{N}}
\toprule
Model & mIoU & R@.3 & R@.5 & R@.7 & gIoU & MAE & CntAcc & Fail \\
\midrule
Audio Flamingo 2 & 0.6 & 0.7 & 0.7 & 0.0 & -53.4 & 25.3 & 56.5 & 0.0 \\
AF2 Sound-CoT & 6.3 & 7.0 & 2.3 & 0.3 & -26.8 & 21.1 & \underline{58.1} & 0.0 \\
Audio Flamingo 3 & 18.7 & 28.6 & 13.6 & 4.7 & -0.5 & 15.6 & \underline{58.1} & 0.0 \\
FireRedAudio & 31.0 & 36.2 & 29.9 & 24.9 & -16.9 & \textbf{5.1} & 36.2 & 40.9 \\
GLM-4-Voice-9B & 5.2 & 6.3 & 2.7 & 0.0 & -42.7 & 135.5 & 51.5 & 7.0 \\
Kimi-Audio-7B & 1.5 & 2.0 & 0.3 & 0.3 & -64.8 & 2397.2 & 42.9 & 21.3 \\
MiDashengLM-7B & 10.9 & 14.3 & 3.0 & 0.0 & -15.8 & 14.7 & 57.5 & 1.0 \\
MiMo-Audio-7B-Instruct & 22.2 & 31.2 & 17.6 & 7.0 & 0.5 & 20.4 & 56.5 & 0.7 \\
MOSS-Audio-4B-Instruct & 30.2 & 38.9 & 20.3 & 15.0 & 10.5 & 16.5 & 57.5 & 0.7 \\
MOSS-Audio-4B-Thinking & \textbf{51.9} & \underline{63.5} & \textbf{50.5} & \textbf{40.9} & \textbf{40.1} & \underline{6.4} & 56.5 & 1.0 \\
MOSS-Audio-8B-Instruct & 26.0 & 31.6 & 18.9 & 13.3 & 3.1 & 11.4 & 57.5 & 1.7 \\
MOSS-Audio-8B-Thinking & \underline{51.9} & \textbf{63.8} & \underline{49.2} & \textbf{40.9} & \underline{40.0} & 18.5 & 56.8 & 0.0 \\
Step-Audio-2-mini-Base & 1.0 & 0.3 & 0.3 & 0.0 & -76.7 & 21.6 & 21.9 & 61.1 \\
Step-Audio-2-mini & 7.8 & 9.0 & 1.0 & 0.0 & -26.3 & 20.5 & 57.5 & 0.7 \\
Step-Audio-2-mini-Think & 2.6 & 2.3 & 0.3 & 0.0 & -45.5 & 21.2 & 54.8 & 5.6 \\
Step-Audio-R1 & 18.7 & 28.9 & 11.6 & 2.7 & -3.7 & 22.8 & 42.2 & 0.0 \\
LLaMA-Omni2-14B & 0.0 & 0.0 & 0.0 & 0.0 & -99.1 & 37.0 & 1.3 & 97.3 \\
\hdashline
LLaMA-3.1-8B-Omni & 0.0 & 0.0 & 0.0 & 0.0 & -99.7 & 11.7 & 0.0 & 99.7 \\
Gemini-3.1-Pro & 46.5 & 60.3 & 46.0 & \underline{31.7} & 36.7 & 8.0 & \textbf{58.3} & 0.0 \\
\hdashline
LAT-Audio & 31.9 & 44.9 & 30.6 & 16.3 & 17.4 & 10.9 & \underline{58.1} & 0.0 \\
TimeAudio & 23.5 & 29.9 & 18.9 & 9.3 & 14.6 & 13.1 & 47.8 & 0.0 \\
\bottomrule
\end{tabular}}
\end{minipage}\par
\vspace{0.08cm}
\hrule
\vspace{0.08cm}
\begin{minipage}[t]{0.48\textwidth}
\raggedright\footnotesize
\textbf{Word.} FireRedAudio leads every overlap and Recall column, reaching 34.5 mIoU and 25.8 R@0.7. The MOSS instruct variants exceed their thinking counterparts at R@0.5 (15.8 versus 8.2 for 4B; 15.1 versus 5.0 for 8B), even though the thinking variants have higher mIoU. This split suggests that the reasoning-style advantage is not uniform at word-level boundary precision.
\end{minipage}\hfill
\begin{minipage}[t]{0.48\textwidth}
\raggedright\footnotesize
\textbf{Acoustic event.} The two MOSS thinking models both reach 51.9 mIoU and 40.9 R@0.7, leading this subset at both average overlap and strict localization; Gemini-3.1-Pro follows at 46.5 mIoU. FireRedAudio's 5.1\,s MAE is lower, but its 40.9\% Fail rate limits that value to 59.1\% coverage, whereas its 31.0 mIoU already charges every failed parse as a miss.
\end{minipage}\par
\vspace{+1cm}
\captionof{table}{Complete per-model results for the acoustic-description and semantic subsets.}
\label{tab:appx2}
\vspace{-0.06cm}
\setlength{\tabcolsep}{2.2pt}
\renewcommand{\arraystretch}{1.2}
\footnotesize
\begin{minipage}[t]{0.49\textwidth}
\vspace{0pt}\centering
\textbf{\strut (a) Acoustic description}\par\vspace{1pt}
\resizebox{\linewidth}{!}{%
\begin{tabular}{l*{8}{N}}
\toprule
Model & mIoU & R@.3 & R@.5 & R@.7 & gIoU & MAE & CntAcc & Fail \\
\midrule
Audio Flamingo 2 & 0.1 & 0.0 & 0.0 & 0.0 & -52.8 & 25.7 & \underline{66.9} & 0.0 \\
AF2 Sound-CoT & 6.7 & 6.9 & 1.5 & 0.4 & -22.9 & 21.0 & 66.5 & 0.0 \\
Audio Flamingo 3 & 15.9 & 20.8 & 13.1 & 4.6 & -2.7 & 15.9 & \underline{66.9} & 0.0 \\
FireRedAudio & 36.8 & 43.8 & 35.8 & 28.8 & 2.8 & \textbf{5.7} & 44.2 & 30.0 \\
GLM-4-Voice-9B & 10.5 & 12.3 & 3.8 & 1.5 & -14.5 & 29.2 & 65.4 & 1.2 \\
Kimi-Audio-7B & 2.0 & 1.2 & 0.0 & 0.0 & -62.7 & 3662.1 & 40.8 & 35.4 \\
MiDashengLM-7B & 11.2 & 16.2 & 3.5 & 1.2 & -15.6 & 17.3 & 64.2 & 3.1 \\
MiMo-Audio-7B-Instruct & 23.7 & 33.5 & 20.0 & 8.1 & 4.3 & 11.9 & 63.8 & 4.2 \\
MOSS-Audio-4B-Instruct & 27.1 & 34.6 & 20.4 & 14.6 & 4.3 & 54.5 & 61.2 & 8.1 \\
MOSS-Audio-4B-Thinking & \textbf{50.6} & \underline{60.0} & \underline{51.9} & \textbf{42.3} & \textbf{41.6} & \underline{6.6} & 64.2 & 1.5 \\
MOSS-Audio-8B-Instruct & 19.1 & 26.2 & 14.6 & 8.5 & -21.8 & 10.2 & 47.3 & 30.8 \\
MOSS-Audio-8B-Thinking & \underline{50.0} & \underline{60.0} & \textbf{52.7} & \textbf{42.3} & 38.9 & 14.1 & 65.8 & 0.0 \\
Step-Audio-2-mini-Base & 2.1 & 1.5 & 1.2 & 0.0 & -66.3 & 65.4 & 30.8 & 55.0 \\
Step-Audio-2-mini & 9.1 & 10.8 & 3.1 & 0.8 & -19.8 & 17.6 & 66.5 & 0.4 \\
Step-Audio-2-mini-Think & 2.6 & 1.9 & 1.2 & 0.4 & -44.0 & 21.3 & 64.2 & 3.8 \\
Step-Audio-R1 & 20.7 & 28.5 & 12.7 & 3.5 & 5.2 & 17.4 & 48.5 & 0.0 \\
LLaMA-Omni2-14B & 0.1 & 0.0 & 0.0 & 0.0 & -98.9 & 40.4 & 1.2 & 98.5 \\
\hdashline
LLaMA-3.1-8B-Omni & 0.1 & 0.0 & 0.0 & 0.0 & -99.3 & 38.4 & 0.8 & 98.8 \\
Gemini-3.1-Pro & 48.1 & \textbf{62.3} & 51.2 & \underline{32.7} & \underline{39.9} & 8.2 & \textbf{67.7} & 0.0 \\
\hdashline
LAT-Audio & 31.3 & 40.8 & 28.5 & 16.5 & 18.2 & 11.2 & \underline{66.9} & 0.0 \\
TimeAudio & 24.3 & 30.8 & 17.7 & 6.5 & 19.4 & 12.9 & 51.5 & 0.0 \\
\bottomrule
\end{tabular}}
\end{minipage}\hfill
\begin{minipage}[t]{0.49\textwidth}
\vspace{0pt}\centering
\textbf{\strut (b) Semantic}\par\vspace{1pt}
\resizebox{\linewidth}{!}{%
\begin{tabular}{l*{8}{N}}
\toprule
Model & mIoU & R@.3 & R@.5 & R@.7 & gIoU & MAE & CntAcc & Fail \\
\midrule
Audio Flamingo 2 & 1.2 & 1.7 & 1.1 & 1.1 & -58.8 & 44.7 & \underline{95.6} & 0.0 \\
AF2 Sound-CoT & 1.8 & 3.3 & 0.0 & 0.0 & -49.7 & 40.4 & \underline{95.6} & 0.0 \\
Audio Flamingo 3 & 7.0 & 10.6 & 2.8 & 1.7 & -27.7 & 28.9 & 95.0 & 0.0 \\
FireRedAudio & \textbf{38.4} & 52.8 & \underline{34.4} & \textbf{21.7} & \underline{29.5} & \underline{10.7} & 91.7 & 2.2 \\
GLM-4-Voice-9B & 4.0 & 5.6 & 1.7 & 0.6 & -61.4 & 970.8 & 60.0 & 36.1 \\
Kimi-Audio-7B & 2.7 & 3.3 & 1.7 & 0.6 & -53.2 & 142.1 & 84.4 & 11.1 \\
MiDashengLM-7B & 5.6 & 6.7 & 1.7 & 0.0 & -41.8 & 62.0 & 81.7 & 12.8 \\
MiMo-Audio-7B-Instruct & 16.0 & 24.4 & 12.8 & 5.0 & -14.9 & 152.0 & 95.0 & 1.1 \\
MOSS-Audio-4B-Instruct & 10.6 & 13.9 & 8.3 & 2.8 & -21.9 & 42.1 & 93.9 & 1.7 \\
MOSS-Audio-4B-Thinking & 33.9 & 47.2 & 31.1 & 14.4 & 24.9 & \textbf{10.4} & \textbf{96.1} & 0.0 \\
MOSS-Audio-8B-Instruct & 11.5 & 15.6 & 8.9 & 4.4 & -21.4 & 36.4 & 92.8 & 2.8 \\
MOSS-Audio-8B-Thinking & \underline{37.4} & \underline{53.9} & \textbf{36.1} & \underline{20.6} & 27.2 & 127.9 & 95.0 & 0.0 \\
Step-Audio-2-mini-Base & 2.2 & 2.2 & 0.6 & 0.6 & -65.0 & 45.1 & 55.0 & 40.6 \\
Step-Audio-2-mini & 7.8 & 10.0 & 4.4 & 1.1 & -22.8 & 77.4 & \underline{95.6} & 0.6 \\
Step-Audio-2-mini-Think & 0.8 & 0.6 & 0.6 & 0.0 & -69.3 & 537.5 & 86.1 & 2.2 \\
Step-Audio-R1 & 13.6 & 17.2 & 7.2 & 3.3 & -12.8 & 97.5 & 67.2 & 0.0 \\
LLaMA-Omni2-14B & 3.2 & 5.0 & 1.1 & 1.1 & -64.8 & 242.5 & 50.0 & 45.0 \\
\hdashline
LLaMA-3.1-8B-Omni & 2.2 & 2.8 & 1.1 & 0.6 & -64.8 & 447.2 & 65.0 & 31.1 \\
Gemini-3.1-Pro & 31.6 & 46.7 & 30.6 & 14.4 & 19.5 & 13.4 & 93.9 & 0.0 \\
\hdashline
LAT-Audio & 35.6 & \textbf{55.0} & 27.8 & 13.9 & \textbf{29.7} & 10.7 & 95.0 & 0.6 \\
TimeAudio & 15.7 & 14.4 & 3.3 & 1.1 & 9.7 & 25.2 & 90.0 & 0.6 \\
\bottomrule
\end{tabular}}
\end{minipage}\par
\vspace{0.08cm}
\hrule
\vspace{0.08cm}
\begin{minipage}[t]{0.48\textwidth}
\raggedright\footnotesize
\textbf{Acoustic description.} The MOSS thinking variants retain nearly their acoustic-event performance on sentence-length descriptions: 50.6/50.0 mIoU here versus 51.9/51.9 on event labels. Gemini-3.1-Pro is close at 48.1 and records the highest R@0.3 (62.3). FireRedAudio has the lowest parsed-only MAE (5.7\,s), but its 30.0\% Fail rate and lower 28.8 R@0.7 show that this conditional error alone does not establish the strongest grounding.
\end{minipage}\hfill
\begin{minipage}[t]{0.48\textwidth}
\raggedright\footnotesize
\textbf{Semantic.} The ranking becomes more heterogeneous: general-purpose FireRedAudio leads mIoU and R@0.7 (38.4 and 21.7), MOSS-Audio-8B-Thinking narrowly leads R@0.3 over FireRedAudio (53.9 versus 52.8), and temporal expert LAT-Audio leads R@0.3 overall and gIoU (55.0 and 29.7). Their leading parsed-only MAEs cluster at 10.4--10.7\,s, so overlap at stricter thresholds, rather than conditional midpoint error, separates their localization profiles.
\end{minipage}\par
\vspace{-0.08cm}
\end{strip}

% Reserve the final appendix page for Tables 8--9 and their analyses.
% NOTE: this final block must NOT use cuted's strip environment. A strip that
% is the last thing before \end{document} has no following body text to trigger
% its output-routine insertion, so some TeX Live versions (incl. arXiv's)
% silently drop the whole page. \onecolumn gives the same full-width layout
% (\textwidth is identical) without that fragility.
\clearpage
\onecolumn
\begingroup
\centering
\vspace{+1cm}
\captionof{table}{Complete per-model results for the rough-semantic and emotion subsets.}
\label{tab:appx3}
\vspace{-0.06cm}
\setlength{\tabcolsep}{2.2pt}
\renewcommand{\arraystretch}{1.2}
\footnotesize
\begin{minipage}[t]{0.49\textwidth}
\vspace{0pt}\centering
\textbf{\strut (a) Rough semantic}\par\vspace{1pt}
\resizebox{\linewidth}{!}{%
\begin{tabular}{l*{8}{N}}
\toprule
Model & mIoU & R@.3 & R@.5 & R@.7 & gIoU & MAE & CntAcc & Fail \\
\midrule
Audio Flamingo 2 & 0.0 & 0.0 & 0.0 & 0.0 & -64.1 & 38.2 & 83.2 & 0.0 \\
AF2 Sound-CoT & 0.8 & 1.2 & 0.0 & 0.0 & -56.3 & 34.9 & 83.9 & 0.0 \\
Audio Flamingo 3 & 7.8 & 8.7 & 3.1 & 0.6 & -31.2 & 23.9 & 83.9 & 0.0 \\
FireRedAudio & \textbf{40.6} & \underline{52.2} & \textbf{39.1} & \textbf{26.7} & \textbf{29.6} & \textbf{5.7} & 82.6 & 0.6 \\
GLM-4-Voice-9B & 2.6 & 3.1 & 1.2 & 0.0 & -62.6 & 433.3 & 51.6 & 34.8 \\
Kimi-Audio-7B & 1.7 & 1.9 & 1.2 & 0.6 & -63.6 & 153.1 & 73.9 & 13.7 \\
MiDashengLM-7B & 3.4 & 4.3 & 0.0 & 0.0 & -53.6 & 32.0 & 66.5 & 20.5 \\
MiMo-Audio-7B-Instruct & 13.5 & 21.1 & 6.2 & 2.5 & -18.0 & 452.1 & 83.9 & 0.6 \\
MOSS-Audio-4B-Instruct & 6.9 & 8.1 & 4.3 & 2.5 & -29.1 & 51.8 & 83.9 & 0.0 \\
MOSS-Audio-4B-Thinking & 25.7 & 37.3 & 21.7 & 7.5 & 9.5 & \underline{8.8} & \textbf{85.1} & 0.0 \\
MOSS-Audio-8B-Instruct & 9.1 & 11.2 & 6.2 & 3.7 & -26.7 & 18.6 & 83.2 & 0.6 \\
MOSS-Audio-8B-Thinking & \underline{36.5} & \textbf{52.8} & \underline{32.3} & \underline{18.6} & \underline{23.5} & 9.5 & \underline{84.5} & 0.0 \\
Step-Audio-2-mini-Base & 1.0 & 1.2 & 1.2 & 0.0 & -73.5 & 55.4 & 46.6 & 44.7 \\
Step-Audio-2-mini & 6.4 & 8.1 & 3.7 & 1.2 & -33.3 & 33.1 & 83.9 & 0.0 \\
Step-Audio-2-mini-Think & 2.9 & 3.7 & 0.6 & 0.6 & -62.4 & 831.7 & 73.9 & 3.7 \\
Step-Audio-R1 & 11.8 & 15.5 & 7.5 & 3.1 & -18.6 & 55.7 & 62.7 & 0.0 \\
LLaMA-Omni2-14B & 4.4 & 5.6 & 3.7 & 0.0 & -54.8 & 118.6 & 63.4 & 21.1 \\
\hdashline
LLaMA-3.1-8B-Omni & 2.4 & 2.5 & 1.2 & 1.2 & -68.6 & 370.6 & 48.4 & 42.2 \\
Gemini-3.1-Pro & 26.8 & 42.2 & 20.5 & 9.3 & 7.9 & 9.0 & 82.6 & 0.0 \\
\hdashline
LAT-Audio & 29.6 & 42.9 & 23.0 & 9.3 & 19.3 & 12.5 & 83.9 & 0.0 \\
TimeAudio & 12.4 & 8.1 & 3.7 & 2.5 & 4.9 & 54.7 & 67.1 & 0.6 \\
\bottomrule
\end{tabular}}
\end{minipage}\hfill
\begin{minipage}[t]{0.49\textwidth}
\vspace{0pt}\centering
\textbf{\strut (b) Emotion}\par\vspace{1pt}
\resizebox{\linewidth}{!}{%
\begin{tabular}{l*{8}{N}}
\toprule
Model & mIoU & R@.3 & R@.5 & R@.7 & gIoU & MAE & CntAcc & Fail \\
\midrule
Audio Flamingo 2 & 0.0 & 0.0 & 0.0 & 0.0 & -78.4 & 48.1 & 64.3 & 0.8 \\
AF2 Sound-CoT & 0.3 & 0.0 & 0.0 & 0.0 & -73.6 & 46.5 & 63.6 & 0.8 \\
Audio Flamingo 3 & 2.5 & 1.6 & 0.8 & 0.0 & -59.2 & 36.4 & 63.6 & 0.0 \\
FireRedAudio & \textbf{25.2} & \textbf{31.8} & \textbf{20.9} & \textbf{10.1} & \underline{-8.4} & \textbf{13.9} & 58.9 & 1.6 \\
GLM-4-Voice-9B & 2.7 & 3.9 & 0.8 & 0.0 & -66.0 & 974.9 & 44.2 & 30.2 \\
Kimi-Audio-7B & 0.4 & 0.8 & 0.0 & 0.0 & -75.7 & 108.0 & 57.4 & 14.7 \\
MiDashengLM-7B & 1.1 & 1.6 & 0.0 & 0.0 & -78.4 & 42.5 & 34.1 & 44.2 \\
MiMo-Audio-7B-Instruct & 7.3 & 10.9 & 3.9 & 1.6 & -44.9 & 39.2 & 65.1 & 0.0 \\
MOSS-Audio-4B-Instruct & 3.1 & 5.4 & 0.8 & 0.0 & -59.2 & 60.6 & 64.3 & 0.8 \\
MOSS-Audio-4B-Thinking & 15.7 & \underline{24.0} & 9.3 & 5.4 & -21.8 & \underline{17.9} & 62.0 & 0.8 \\
MOSS-Audio-8B-Instruct & 4.3 & 6.2 & 1.6 & 0.0 & -59.0 & 63.9 & 62.8 & 4.7 \\
MOSS-Audio-8B-Thinking & \underline{18.2} & \underline{24.0} & \underline{12.4} & \underline{7.8} & -15.9 & 132.7 & \underline{65.9} & 0.0 \\
Step-Audio-2-mini-Base & 0.7 & 1.6 & 0.8 & 0.0 & -82.6 & 66.4 & 35.7 & 43.4 \\
Step-Audio-2-mini & 2.5 & 2.3 & 0.8 & 0.0 & -58.5 & 30.5 & 65.1 & 0.0 \\
Step-Audio-2-mini-Think & 1.5 & 1.6 & 1.6 & 0.8 & -76.4 & 413.2 & 55.8 & 4.7 \\
Step-Audio-R1 & 4.0 & 3.1 & 0.8 & 0.0 & -42.0 & 76.2 & 48.1 & 0.0 \\
LLaMA-Omni2-14B & 1.7 & 0.0 & 0.0 & 0.0 & -60.3 & 65.3 & 56.6 & 8.5 \\
\hdashline
LLaMA-3.1-8B-Omni & 0.6 & 0.0 & 0.0 & 0.0 & -70.8 & 77.8 & 46.5 & 27.1 \\
Gemini-3.1-Pro & 11.4 & 11.6 & 4.7 & 3.1 & -26.9 & 20.4 & \textbf{66.7} & 0.0 \\
\hdashline
LAT-Audio & 13.4 & 17.1 & 7.0 & 1.6 & -13.9 & 21.2 & 64.3 & 0.8 \\
TimeAudio & 9.0 & 6.2 & 1.6 & 0.0 & \textbf{-0.1} & 32.5 & 62.0 & 0.0 \\
\bottomrule
\end{tabular}}
\end{minipage}\par
\vspace{0.08cm}
\hrule
\vspace{0.08cm}
\begin{minipage}[t]{0.48\textwidth}
\raggedright\footnotesize
\textbf{Rough semantic.} FireRedAudio leads mIoU, R@0.5, R@0.7, and gIoU (40.6, 39.1, 26.7, and 29.6), with only 0.6\% Fail. MOSS-Audio-8B-Thinking remains close to its exact-semantic mIoU (36.5 versus 37.4), whereas its 8B instruct counterpart is much lower at 9.1. Although the subsets come from different sources, this descriptive pattern is consistent with the thinking variant being more robust to underspecified wording.
\end{minipage}\hfill
\begin{minipage}[t]{0.48\textwidth}
\raggedright\footnotesize
\textbf{Emotion.} FireRedAudio leads with 25.2 mIoU and 10.1 R@0.7, followed by MOSS-Audio-8B-Thinking at 18.2 and 7.8. The best strict Recall is still the lowest among the short-audio subsets, consistent with emotional episodes having less sharply defined boundaries than discrete events. TimeAudio instead has the best gIoU ($-0.1$) but only 9.0 mIoU, showing that its temporal-error profile does not translate into the strongest interval overlap.
\end{minipage}\par
\vspace{1cm}
\captionof{table}{Complete per-model results for the two long-audio subsets.}
\label{tab:appx4}
\vspace{-0.06cm}
\setlength{\tabcolsep}{2.2pt}
\renewcommand{\arraystretch}{1.2}
\footnotesize
\begin{minipage}[t]{0.49\textwidth}
\vspace{0pt}\centering
\textbf{\strut (a) Long -- semantic}\par\vspace{1pt}
\resizebox{\linewidth}{!}{%
\begin{tabular}{l*{8}{N}}
\toprule
Model & mIoU & R@.3 & R@.5 & R@.7 & gIoU & MAE & CntAcc & Fail \\
\midrule
Audio Flamingo 2 & 0.0 & 0.0 & 0.0 & 0.0 & -93.8 & 506.1 & \textbf{97.2} & 0.0 \\
AF2 Sound-CoT & 0.1 & 0.0 & 0.0 & 0.0 & -92.3 & 501.4 & \textbf{97.2} & 0.0 \\
Audio Flamingo 3 & 0.2 & 0.0 & 0.0 & 0.0 & -91.1 & 540.7 & \underline{96.7} & 0.6 \\
FireRedAudio & \textbf{20.6} & \textbf{26.7} & \textbf{16.1} & \textbf{7.8} & \textbf{2.4} & \textbf{33.7} & 92.2 & 0.6 \\
GLM-4-Voice-9B & 0.0 & 0.0 & 0.0 & 0.0 & -99.0 & 570.9 & 8.9 & 91.1 \\
Kimi-Audio-7B & 0.7 & 0.6 & 0.6 & 0.0 & -85.8 & 737.7 & 63.3 & 33.9 \\
MiDashengLM-7B & 0.1 & 0.0 & 0.0 & 0.0 & -95.2 & 488.1 & 42.2 & 55.6 \\
MiMo-Audio-7B-Instruct & 4.2 & 5.6 & 2.8 & 0.6 & -55.8 & 212.7 & \underline{96.7} & 0.6 \\
MOSS-Audio-4B-Instruct & 0.4 & 0.0 & 0.0 & 0.0 & -84.9 & 1826.0 & \underline{96.7} & 1.1 \\
MOSS-Audio-4B-Thinking & 3.2 & 4.4 & 3.9 & 0.6 & -72.4 & 320.4 & 93.9 & 2.8 \\
MOSS-Audio-8B-Instruct & 1.6 & 2.2 & 0.6 & 0.0 & -76.4 & 842.8 & 95.0 & 2.2 \\
MOSS-Audio-8B-Thinking & 4.8 & 7.2 & 3.9 & 0.6 & -61.0 & 3739.1 & 95.0 & 0.0 \\
Step-Audio-2-mini-Base & 0.0 & 0.0 & 0.0 & 0.0 & -99.9 & 1230.3 & 2.8 & 97.2 \\
Step-Audio-2-mini & 0.0 & 0.0 & 0.0 & 0.0 & -98.6 & 531.3 & 3.9 & 96.1 \\
Step-Audio-2-mini-Think & 0.0 & 0.0 & 0.0 & 0.0 & -98.9 & 469.0 & 6.1 & 93.3 \\
Step-Audio-R1 & 1.3 & 1.7 & 0.6 & 0.6 & -78.0 & 1990.2 & 64.4 & 0.0 \\
LLaMA-Omni2-14B & 0.5 & 0.0 & 0.0 & 0.0 & -90.4 & 459.8 & 36.7 & 62.2 \\
\hdashline
LLaMA-3.1-8B-Omni & 0.4 & 0.6 & 0.0 & 0.0 & -89.6 & 755.0 & 56.7 & 40.0 \\
Gemini-3.1-Pro & \underline{12.9} & \underline{16.1} & \underline{7.2} & \underline{2.2} & \underline{-19.5} & 119.4 & 95.0 & 1.1 \\
\hdashline
LAT-Audio & 8.6 & 11.7 & 5.6 & 0.6 & -29.7 & \underline{95.2} & \textbf{97.2} & 0.0 \\
TimeAudio & 1.6 & 1.7 & 0.6 & 0.0 & -69.0 & 460.7 & 71.7 & 13.9 \\
\bottomrule
\end{tabular}}
\end{minipage}\hfill
\begin{minipage}[t]{0.49\textwidth}
\vspace{0pt}\centering
\textbf{\strut (b) Long -- acoustic description}\par\vspace{1pt}
\resizebox{\linewidth}{!}{%
\begin{tabular}{l*{8}{N}}
\toprule
Model & mIoU & R@.3 & R@.5 & R@.7 & gIoU & MAE & CntAcc & Fail \\
\midrule
Audio Flamingo 2 & 0.0 & 0.0 & 0.0 & 0.0 & -93.7 & 519.6 & \underline{66.2} & 0.0 \\
AF2 Sound-CoT & 0.3 & 0.8 & 0.0 & 0.0 & -89.9 & 514.6 & \textbf{66.9} & 0.0 \\
Audio Flamingo 3 & 0.7 & 1.2 & 0.4 & 0.4 & -87.4 & 509.2 & \textbf{66.9} & 0.0 \\
FireRedAudio & \textbf{21.7} & \textbf{27.3} & \textbf{21.5} & \textbf{17.3} & \textbf{-33.6} & \textbf{185.4} & 50.4 & 20.8 \\
GLM-4-Voice-9B & 0.0 & 0.0 & 0.0 & 0.0 & -98.5 & 507.5 & 7.3 & 86.2 \\
Kimi-Audio-7B & 0.2 & 0.0 & 0.0 & 0.0 & -90.8 & 944.0 & 22.3 & 66.5 \\
MiDashengLM-7B & 0.0 & 0.0 & 0.0 & 0.0 & -96.9 & 557.8 & 19.2 & 71.9 \\
MiMo-Audio-7B-Instruct & 2.8 & 5.0 & 1.5 & 0.4 & -79.0 & 1920.8 & 63.1 & 6.5 \\
MOSS-Audio-4B-Instruct & 0.2 & 0.0 & 0.0 & 0.0 & -91.0 & 511.8 & 57.3 & 12.3 \\
MOSS-Audio-4B-Thinking & 1.2 & 0.8 & 0.4 & 0.4 & -78.4 & 340.8 & 65.8 & 1.2 \\
MOSS-Audio-8B-Instruct & 0.6 & 0.8 & 0.4 & 0.4 & -87.8 & 462.4 & 42.7 & 36.5 \\
MOSS-Audio-8B-Thinking & 4.1 & 5.0 & 3.1 & 2.7 & -66.4 & 3438.4 & 64.2 & 0.4 \\
Step-Audio-2-mini-Base & 0.0 & 0.0 & 0.0 & 0.0 & -100.0 & 672.3 & 1.2 & 98.8 \\
Step-Audio-2-mini & 0.0 & 0.0 & 0.0 & 0.0 & -99.7 & 652.7 & 4.2 & 93.8 \\
Step-Audio-2-mini-Think & 0.0 & 0.0 & 0.0 & 0.0 & -100.0 & -- & 0.0 & 100.0 \\
Step-Audio-R1 & 1.6 & 1.5 & 1.5 & 0.4 & -82.0 & 573.5 & 55.0 & 0.8 \\
LLaMA-Omni2-14B & 0.0 & 0.0 & 0.0 & 0.0 & -99.9 & 559.8 & 0.8 & 99.2 \\
\hdashline
LLaMA-3.1-8B-Omni & 0.0 & 0.0 & 0.0 & 0.0 & -99.9 & 488.5 & 1.5 & 98.1 \\
Gemini-3.1-Pro & \underline{7.2} & \underline{9.2} & \underline{5.0} & 2.3 & -65.6 & 286.5 & 41.2 & 0.0 \\
\hdashline
LAT-Audio & 5.5 & 7.3 & 4.2 & \underline{3.8} & \underline{-61.7} & \underline{265.5} & \underline{66.2} & 0.8 \\
TimeAudio & 0.6 & 0.8 & 0.0 & 0.0 & -82.4 & 489.5 & 46.2 & 3.8 \\
\bottomrule
\end{tabular}}
\end{minipage}\par
\vspace{0.08cm}
\hrule
\vspace{0.08cm}
\begin{minipage}[t]{0.48\textwidth}
\raggedright\footnotesize
\textbf{Long -- semantic.} FireRedAudio leads at 20.6 mIoU and is the only model with positive gIoU (2.4); omni-modal Gemini-3.1-Pro and temporal expert LAT-Audio follow at 12.9 and 8.6. Every other general-purpose system remains at or below 4.8 mIoU. FireRedAudio's 33.7\,s parsed-only MAE also has 99.4\% coverage, so both its conditional and full-coverage metrics support the same ranking on this subset.
\end{minipage}\hfill
\begin{minipage}[t]{0.48\textwidth}
\raggedright\footnotesize
\textbf{Long -- acoustic description.} FireRedAudio leads with 21.7 mIoU and 17.3 R@0.7; Gemini-3.1-Pro is second in mIoU at 7.2, whereas LAT-Audio is second at the strict threshold with 3.8 R@0.7. FireRedAudio's 185.4\,s MAE covers only 79.2\% of samples because Fail is 20.8\%. Its leading mIoU and Recall nevertheless include those failures as misses, so the long-form advantage is not created by MAE filtering.
\end{minipage}\par
\vspace{0.08cm}
\begin{minipage}[t]{0.98\textwidth}
\raggedright\footnotesize
% \textbf{Cross-subset reading.} Short-audio leadership does not transfer automatically to long recordings: the MOSS thinking variants approach 50--52 mIoU on the two short acoustic subsets but reach only 3.2--4.8 on long semantic and 1.2--4.1 on long acoustic-description queries. Parsing reliability is a separate axis as well. MOSS-Audio-8B-Thinking parses 100\% and 99.6\% of the two long subsets, yet its parsed-only MAEs exceed 3{,}400\,s and its mIoUs remain below 5. Conversely, FireRedAudio's Fail rate is at most 2.2\% on five subsets but rises on all three event/description subsets (40.9\%, 30.0\%, and 20.8\%), concentrating its format-reliability weakness in these query categories even though it retains the best long-form overlap. The best long-audio mIoU is still only 21.7, leaving a large gap to reliable long-form grounding.
\end{minipage}\par
\vspace{-0.08cm}
\endgroup

\end{document}